\documentclass[11pt]{article}

\usepackage[utf8]{inputenc}
\usepackage{mathpazo}
\usepackage{bm}
\usepackage{aurical}
\usepackage{import}
\usepackage[toc,page]{appendix}\usepackage{latexsym,amsfonts,amsmath,
amssymb,mathrsfs,bbold,mathtools,esint,amsthm,mathtools}
\usepackage{dsfont}
\usepackage{mathtools}
\usepackage{float}
\usepackage{graphicx}
\usepackage{cite}
\usepackage{hyperref}
\usepackage{setspace}
\usepackage{color}
\usepackage{slashed}
\usepackage{cleveref}
\numberwithin{equation}{section}
\usepackage[colorinlistoftodos]{todonotes}
\usepackage[affil-it]{authblk}
\usepackage{float}
\usepackage{indentfirst}
\usepackage{soul}
\usepackage{booktabs}
\usepackage{eso-pic,graphicx}
\usepackage{nicefrac}
\usepackage{epsfig}

\newcommand{\sn}{\mathrm{sn}}
\newcommand{\dn}{\mathrm{dn}}
\newcommand{\cn}{\mathrm{cn}}

\DeclareMathOperator\perm{perm}
\DeclareMathOperator\sign{sign}
\DeclareMathOperator\supp{supp}
\DeclareMathOperator\sech{sech}

\usepackage[a4paper]{geometry}
\usepackage{a4wide}

\usepackage{color,hyperref}
\definecolor{darkblue}{rgb}{0.0,0.0,0.3}
\hypersetup{colorlinks,breaklinks,
            linkcolor=darkblue,urlcolor=darkblue,
            anchorcolor=darkblue,citecolor=darkblue}

\title{Exceptional solutions to the eight-vertex model and integrability of anisotropic extensions of massive fermionic models}
\author[1]{A. Melikyan\footnote{\href{mailto:amelik@gmail.com}{amelik@gmail.com}}}
\author[2]{G. Weber\footnote{\href{mailto:gabrielweber@usp.br}{gabrielweber@usp.br}} }
\affil[1]{Instituto de Física\\
Universidade de Brasília\\
70910-900, Brasília, DF, Brasil}
\affil[2]{Escola de Engenharia de Lorena\\
Universidade de São Paulo\\
12602-810, Lorena, SP, Brasil}

\usepackage{palatino}

\begin{document}
\maketitle

\begin{abstract}
We consider several anisotropic extensions of the Belavin model, and show that integrability holds also for the massive case for some specific relations between the coupling constants. This is done by relating the S-matrix factorization property to the exceptional solutions of the eight-vertex model. The relation of exceptional solutions to the XXZ and six-vertex models is also shown.
\end{abstract}

\paragraph{Keywords:} Exactly Solvable Models, Bethe Ansatz; Continuum models;
Integration of Completely integrable systems by inverse spectral and scattering methods; Quantum Field Theory.

\section{Introduction}
\label{intro}
Integrable properties of the string theory on the $AdS_5 \times S^5$ background have prompted a renewed interest in subtle properties of classical and quantum integrable systems \cite{Beisert:2010jr,Bombardelli:2016rwb}. One such subtle issue is the non-ultralocality of the algebra of Lax connections. Besides string theory, it plagues many important integrable models \cite{Freidel:1991jv,Freidel:1991jx,Faddeev:1985qu,Maillet:1985ec,Maillet:1985ek,Maillet:1985fn}. So far there are no reliable general methods to quantize such systems, despite many interesting directions and results \cite{Delduc:2012qb,Delduc:2012vq,Kundu:2003cu,SemenovTianShansky:1995ha,Kundu:1996hb,Melikyan:2016gkd,Melikyan:2014yma,Schmidtt2017,Appadu:2017td,Vicedo:2017cge,Appadu:2018ioy,Schmidtt:2018hop}. Another difficulty, also shared by strings, is the singular nature of the interaction terms in the quantum-mechanical picture. Two characteristic examples of such models are the Landau-Lifshitz ($LL$) \cite{Faddeev:1987ph,Klose:2006dd,Melikyan:2008ab,Melikyan:2008cy,Melikyan:2010bi,Melikyan:2010fr,Roiban:2006yc,Stefanski:2007dp,Tirziu:2006ve} and the Alday-Arutyunov-Frolov ($AAF$) models \cite{Alday:2005jm,Arutyunov:2005hd,Melikyan:2011uf,Melikyan:2012kj,Melikyan:2014yma,Melikyan:2014mfa}, arising in the $su(2)$ and $su(1 \vert 1)$ sectors of string theory on $AdS_5 \times S^5$ respectively. Some initial steps to deal with these issues have already been considered in \cite{Melikyan:2008ab,Melikyan:2010fr,Melikyan:2012kj,Melikyan:2014yma,Melikyan:2016gkd}. 

The $LL$ model is a classical integrable model and has been extensively investigated using various methods \cite{Faddeev:1987ph,Klose:2006dd}. It is the  continuous limit of the Heisenberg spin chain, which is much simpler to deal with due to absence of singular interactions. It is, however, an interesting toy model to quantize directly in the continuous limit, \emph{i.e.}, without invoking its discretized version \cite{Sklyanin:1988}. This activity becomes particularly relevant, because the lattice regularization of the $AAF$ model has not so far been constructed, and, therefore, it is an important open problem to understand the quantization of such continuous systems. Due to the singular nature of the interaction terms in the Lagrangians of both the $LL$ and $AAF$ models, it has been shown that a consistent quantization procedure requires the construction of self-adjoint extensions of the Hamiltonians of both theories \cite{Melikyan:2008ab,Melikyan:2008cy,Melikyan:2010fr,Melikyan:2014mfa,Melikyan:2016gkd}. The reason is the presence of terms proportional to derivatives of the fields in the interaction Lagrangians, which results in terms containing derivatives of delta-function in the quantum mechanical Hamiltonians.\footnote{For details see, for example, \cite{Gitman:2012ru}} In addition, the $AAF$ model is also non-ultralocal, which further complicates the analysis of its integrable properties \cite{deVega:1983gy,Maillet:1985ek,Melikyan:2012kj,Melikyan:2014yma}. To properly address some of these problems, one has to consider the fields as operator-valued distributions, and use a suitable regularized product of quantum fields. This has been shown to produce  consistent and known results for the $LL$ model \cite{Melikyan:2010fr}, and, for the $AAF$ model, it has correctly reproduced the free fermion limit \cite{Melikyan:2016gkd}. In addition, it was shown that employing distributions and a regularized field product in the quantum theory naturally reproduces Maillet symmetrization procedure in the classical limit \cite{Melikyan:2016gkd}. Nonetheless, it is still very hard to proceed with concrete calculations in the case of the $AAF$ model,\footnote{The explicit form of the Lagrangian for the $AAF$ model is given in \eqref{ext:aaf_lagrangian}.} since the interaction terms go up to the sixth order in the fermion fields and their derivatives. 

The $LL$ model has another interesting feature which helps to understand its integrable properties without appealing to its discretized version. Namely, it is the low-energy limit of the Faddeev-Reshetikhin ($FR$) model \cite{Faddeev:1985qu,Das:2007tb,Delduc:2012qb,Delduc:2012mk,Delduc:2012vq,Delduc:2014uaa} - integrating out the high-energy modes of the $FR$ model in the $QFT$ description naturally leads to the derivatives of the fields in the interaction terms of the $LL$ model \cite{Klose:2006dd}. The main advantage of the $FR$ model is that the interaction term needed to extract the two-particle $S$-matrix does not contain any derivative of the fields, and, therefore, there are no singularities in the corresponding quantum-mechanical Hamiltonian term.  Besides the method proposed by Fadeev and Reshetikhin, one can also solve the model by fermionization techiques \cite{Polyakov:1983tt,Polyakov:1984et}. In the case of the $AAF$ model, however, there is no such simpler high-energy theory from which it could be reproduced in the low-energy limit. In fact, there are not many known purely fermionic continuous integrable systems to begin with. The most well-known Thirring model does not clearly contain enough fermionic degrees of freedom in order to integrate out the high-energy modes and still obtain as a result a non-trivial fermionic model similar to the $AAF$ model. Thus, motivated by these considerations, the main goal of this paper is to find a new class of purely fermionic integrable models without derivatives of the fields in the interaction terms, and with enough fermionic degrees of freedom so that the resulting model in the low-energy limit is still non-trivial. In addition, we require that such a fermionic model be massive and contain interaction terms similar to those of the $AAF$ model. 

The simplest known candidate satisfying the first requirement is the model originally considered by Belavin \cite{Belavin:1979pq} (see also \cite{Vigman:1977la,Vaks:1961vl}) and subsequently explored in \cite{Melzer:1995vw,Reshetikhin:1985ru,Truong:1981fi,Gerdjikov:2012ga,Wang:2000nd,Kirillov:1987jp,Ivanov:1986kt,Marino:1986ji,Doria:1984ym,Dutyshev:1980vn}, which exhibits asymptotic freedom and dimensional transmutation. It is a $su(2)$ symmetric four-component fermionic model which has been shown to be integrable in the zero-mass limit. Our first result is that Belavin's model is in fact integrable also in the massive case for some particular cases of the coupling constants. The analysis requires several steps. First, as we discussed above, the fields should be considered as operator-valued distributions and, furthermore, the operator products should be regularized to avoid singular terms. Next, the integrability in the quantum case reduces to an analysis similar to that of the $R$-matrix factorization condition for the $XYZ$ model conducted by Baxter \cite{Baxter:1982zz}. Recall, that Baxter has deduced a general condition for such factorization to occur, and has given a general solution in terms of elliptic functions.\footnote{For a recent overview see the monograph \cite{Samaj:2013yva} and the references therein.} Indeed, as shown by Belavin \cite{Belavin:1979pq} and later by Dutyshev \cite{Dutyshev:1980vn}, it can happen only for the massless case. There are, however, some exceptional cases for which the $R$-matrix factorization may also occur, corresponding to the situations in which Baxter's condition is not valid due to the degeneracy of the resulting Yang-Baxter equations. Such exceptional cases have been classified and discussed in details in \cite{Khachatryan:2012wy} (see also \cite{Hietarinta:1991ti,Gomez:1996az}). For Belavin's model, we find one such exceptional solution for some particular relation between the coupling constants.\footnote{We note here, that such a condition between the coupling constants also happens in the $AAF$ model, for which the $S$-matrix factorization occurs provided constraint on the two coupling constants of the theory \cite{Melikyan:2011uf}.}

In addition, we generalize our analysis by adding two new interaction terms that mimic the interaction terms of the $AAF$ model. The resulting $S$-matrices are shown not to have the form of the $R$-matrix in the $XYZ$ model, but rather its generalized inhomogeneous form, which had been considered in \cite{Khachatryan:2012wy}, and where its various non-trivial solutions have been classified, including the exceptional cases. Here too, we find that there are such exceptional solutions, corresponding to the massive theories. 

Our paper is organized as follows. In section \ref{section:Belavin}, we consider the anisotropic Belavin model, introduce Sklyanin's product to deal with singular operator products, and obtain the conditions necessary for the diagonalization of the two-particle sector. In section \ref{section:Extensions}, we extend these considerations for the Lagrangian with new interaction terms resembling those of the $AAF$ model. In section \ref{section:8vertex}, we consider the necessary conditions for the integrability of the model, \emph{i.e.}, the $S$-matrix factorization condition, and show that it reduces to finding the exceptional solutions of the eight-vertex model with homogeneous and inhomogeneous $S$-matrix structures. In section \ref{section:xxz}, we show the relation of our exceptional solutions to the $XXZ$ and six-vertex models. In the last section \ref{section:Conclusion}, we outline some possible continuations and open problems.

\section{Anisotropic Belavin model}
\label{section:Belavin}
We start from the classical Lagrangian density 
\begin{align}\label{DM_Original_Lagrangian}
	\mathscr{L} = i \bar{\psi} \gamma_{\mu} \partial^{\mu} \psi - m \bar{\psi} \psi - \frac{g_0}{2} \left( \bar{\psi} \gamma^{\mu} \psi \right)  \left( \bar{\psi} \gamma_{\mu} \psi \right) -\frac{g_a}{2}  \left( \bar{\psi} \gamma^{\mu} \tau^a \psi \right) \left( \bar{\psi} \gamma_{\mu} \tau^a \psi \right)
\end{align}
introduced by Belavin in \cite{Belavin:1979pq} and generalized to the anisotropic case in \cite{Dutyshev:1980vn} to describe the interacting theory of an isotopic fermion in $(1+1)$-dimensions $\psi_i^{\alpha}(x,t)$ with spinorial index $i \in \{1,2\}$ and isotopic index $\alpha \in \{1,2\}$. The isotropic interaction coupling constant $g_0$ is supplemented by three distinct coupling constants $g_a$, $a \in \{1,2,3\}$ for the isotopic interaction. The $\gamma$- and $\tau$- matrices are chosen as
\begin{align}
	&\gamma^0 = \sigma_x, \quad \gamma^1 = i \sigma_y, \quad \text{and} \quad \gamma^3 = \gamma^0 \gamma^1 = - \sigma_z ,\\
	&\tau^1 = \sigma_x, \quad \tau^2 = \sigma_y, \quad \text{and} \quad \tau^3 = \sigma_z,
\end{align}
in terms of the usual representation for the Pauli matrices:
\begin{align}
	\sigma_x = \begin{bmatrix} 0&1\\ 1&0\\ \end{bmatrix}, \quad \sigma_y = \begin{bmatrix} 0&-i\\ i&0\\ \end{bmatrix}, \quad \text{and} \quad \sigma_z = \begin{bmatrix} 1 & 0\\ 0 &-1\\ \end{bmatrix}.
\end{align}
The spacetime metric is $g_{\mu \nu} = \text{diag} (1,-1)$ and the metric in the isotopic space is simply $\delta_{ab}$. 

The Hamiltonian density corresponding to \eqref{DM_Original_Lagrangian} is
\begin{align} \label{DM_Original_Hamiltonian}
	\mathscr{H} = i {\psi^{\dagger}}^{\alpha}_j \sigma_z^{jk} \partial_1 \psi^{\alpha}_k + m {\psi^{\dagger}}^{\alpha}_j \sigma_x^{jk} \psi^{\alpha}_k - \frac{1}{2} {\psi^{\dagger}}^{\alpha}_j{\psi^{\dagger}}^{\beta}_l G_{\alpha \beta, {\alpha'} {\beta'}} \Lambda^{jl,km} \psi^{{\alpha'}}_k \psi^{{\beta'}}_m,
\end{align}
where we introduced the following matrices 
\begin{align}
	G_{\alpha \beta, {\alpha'} {\beta'}} &= g_0 \delta_{\alpha {\alpha'}} \delta_{\beta, {\beta'}} + g_a \tau^a_{\alpha {\alpha'}} \tau^a_{\beta {\beta'}} = \left( g_0 \mathbb{1} \otimes \mathbb{1}  + g_a \tau^a \otimes \tau^a \right)_{\alpha \beta, {\alpha'}{\beta'}},\label{DM:G_tensor_orig}\\
	 \Lambda^{jl,km}  &= \delta^{jk} \delta^{lm} - \sigma_z^{jk} \sigma_z^{lm} = \left( \mathbb{1} \otimes \mathbb{1} - \sigma_z \otimes \sigma_z \right)^{jl,km}\label{DM:Lambda_tensor_orig}
\end{align}
for the interaction terms. The Poisson algebra reads:
\begin{align}\label{DM_PB}
	\left\{ \psi^{\alpha}_j (x), {\psi^{\dagger}}_k^{\beta}(y) \right\} = i \delta^{\alpha \beta} \delta_{jk} \delta(x-y).
\end{align}

\subsection{Quantization and Sklyanin product}
\label{Quantization}
To properly quantize and diagonalize a continuous hamiltonian such as \eqref{DM_Original_Hamiltonian} without resorting to a lattice regularization, it is necessary to deal with the singularities associated with operator products at the same point. As discussed in \cite{Melikyan:2016gkd}, this requires a two-step regularization procedure involving both the regularization of the fields and the operator product. 

First, one has to take care of the singularities related to the canonical anticommutation relations:
\begin{align}\label{DM_CAR}
	\left\{ \psi^{\alpha}_j (x), {\psi^{\dagger}}_k^{\beta}(y) \right\} =  \delta^{\alpha \beta} \delta_{jk} \delta(x-y),
\end{align}
in the limit $x\to y$. One solution to this problem is to formulate the quantum theory in terms of operator valued distributions
\begin{align}\label{DM:OVD}
	{\psi_{\scriptscriptstyle\mathscr{F}}}_k^{\alpha} (x) = \int dz \: \mathscr{F}_{\mu}(x,z) \psi^{\alpha}_k(z),
\end{align}
in which the function $\mathscr{F}_{\mu}(x,y)$ is a symmetric test function which satisfies
\begin{align}\label{DM:F_limit}
	\mathscr{F}_{\mu}(x,y) \xrightarrow{\mu \to 0} \zeta \delta(x-y).
\end{align}
The $c$-number $\zeta$ can be subsequently fixed to 
\begin{align}\label{DM:c-number}
	\zeta^2 = \frac{\zeta_0}{\delta_{\mu}(0)},
\end{align}
where $\zeta_0 \in \mathbb{R}$, by requiring the algebra of regularized fields to be finite in the limit $x \to y$.  In \eqref{DM:c-number}, the function $\delta_{\mu}$ corresponds to the following regularization of the delta function
\begin{align}\label{DM:delta_reg}
	\delta_L (x) =  \frac{1}{2\pi} \int^{\nicefrac{L}{2}}_{-\nicefrac{L}{2}} dz \: e^{i x z},
\end{align}
with respect to the regularizing parameter $L=\nicefrac{1}{\mu}$ associated to the length of the box in which we consider our system, resulting, as a consequence, in asymptotic Bethe Ansatz equations.

It is important to emphasize that treating the fields as operator valued distributions has fundamental consequences in the context of integrable models. For instance, it is in general only possible to obtain a meaningful Yang-Baxter equation from which one can subsequently derive the trace identities in terms of fields regularized as \eqref{DM:OVD}. One notable example of this situation is given by the $LL$ model, in which the afore mentioned procedure allows to avoid ill-defined expressions such as $\partial_x^2 \delta(0)$. Furthermore, only by treating the fields as operator valued distributions one can construct the appropriate self-adjoint extensions and derive the corresponding spectra \cite{Melikyan:2010fr}.

There remains, however, the singularities arising from possible discontinuities of the wave-function and its derivatives during the diagonalization of the quantum Hamiltonian. This issue can be sorted out by also regularizing the operator product of fields. Let $A_i$, $i \in \mathbb{Z}_+$ denote an arbitrary set of fields and consider the operator corresponding to \eqref{DM:OVD}, \emph{i.e.},
\begin{align}\label{DM:generic_field}
	{A_{\scriptscriptstyle\mathscr{F}}}_i (x) = \int dz_i \: \mathscr{F}_{\mu_i}(z_i,x) A_i(z_1).
\end{align}
We define the $\circ$-product as \cite{Melikyan:2016gkd}:\footnote{The $\circ$-product is a modified \emph{Sklyanin product} of \cite{Sklyanin:1988}.}
\begin{align}\label{DM:sklyanin_prod_general}
A_{1}(x) \circ \ldots  \circ A_{k}(x) &\equiv \fint_x  d\xi_{1} \ldots d\xi_{k} \;  A_{1}(\xi_{1}) \cdot \cdots \cdot A_{k}(\xi_{k}) \\
&:=  \lim_{\Delta V_i \to 0}\frac{1}{\sum_{i=1}^{k!} V_i} \sum\limits_{i=1}^{k!}\int\limits_{\Delta V_{i}} d\xi_{1} \ldots d\xi_{k} \;  A_{1}(\xi_{1}) \cdot \cdots \cdot A_{k}(\xi_{k}). \nonumber
\end{align}
Here the integration is taken over a $k$-dimensional hypercube of side $\Delta \in \mathbb{R}$ around the point $x$ where all possible singular $(k-1)$-dimensional hyperplanes are removed from the integration domain. Hence, there are $k!$ disjoint regions of volume $\Delta V_i$, $1 \leq i \leq k!$ corresponding to all possible orderings of the variables $\zeta_j$, $1\leq j \leq k$ separated by the length of a regularization parameter $\epsilon \in \mathbb{R}$. As argued in \cite{Melikyan:2016gkd}, the $\circ$-product \eqref{DM:sklyanin_prod_general} correctly reproduces the Maillet symmetrization procedure for non-ultralocal models \cite{Maillet:1985ek} in the classical limit. To further clarify the meaning of the $\circ$-product \eqref{DM:sklyanin_prod_general} and prepare for the diagonalization of the quantum hamiltonian, in the following, we evaluate in details the $\circ$-product of two and four $\mathscr{F}$-regularized fields \eqref{DM:generic_field}.

The $\circ$-product of two operator valued distributions corresponds to:
\begin{align}\label{DM:2f_product}
	P^{(2)} &= {A_{\scriptscriptstyle\mathscr{F}}}_1(x) \circ {A_{\scriptscriptstyle\mathscr{F}}}_2(x) =  \int dz_1 dz_2 \:  \fint_x d \xi_1 d\xi_2 \:\mathscr{F}_{\mu_1}(z_1,\xi_1) \mathscr{F}_{\mu_2}(z_2, \xi_2) A_1(z_1) A_2(z_2) \\
	=& \lim_{\Delta \to \epsilon} \frac{1}{(\Delta - \epsilon)^2} \int dz_1 dz_2 \left\{ \int_{x-\frac{\Delta}{2}}^{x+\frac{\Delta}{2}-\epsilon} d \xi_2 \int_{\xi_2+\epsilon}^{x+\frac{\Delta}{2}} d \xi_1 \mathscr{F}_{\mu_1}(z_1,\xi_1) \mathscr{F}_{\mu_2}(z_2,\xi_2) A_1(z_1) A_2(z_2) \right. \nonumber \\
	&+ \left. \int_{x-\frac{\Delta}{2}}^{x+\frac{\Delta}{2}-\epsilon} d \xi_1 \int_{\xi_1+\epsilon}^{x+\frac{\Delta}{2}} d \xi_2 \: \mathscr{F}_{\mu_1}(z_1,\xi_1) \mathscr{F}_{\mu_2} (z_2,\xi_2) A_1(z_1) A_2(z_2) \right\} \nonumber \\
	\equiv& P^{(2)}_a + P^{(2)}_b. \nonumber
\end{align}
In \eqref{DM:2f_product}, the integration is taken over a square of side $\Delta$ minus a strip of size $2 \epsilon$ around the diagonal $\xi_1 = \xi_2$, which corresponds to the union of the triangles above the line $\xi_1 = \xi_2 - \epsilon$ and bellow the line $\xi_1 = \xi_2 + \epsilon$. Hence, $\Delta V_i = \frac{1}{2} \left( \Delta - \epsilon \right)^2$, $i=1,2$ so that the limit $\Delta V_i \to 0$ becomes $\Delta \to \epsilon$. 

To compute the first integral in \eqref{DM:2f_product}, we first note that the integrand is a product of continuous functions. Thus, we can perform the integrations over $\xi_1$ and $\xi_2$ by simply invoking the mean value theorem to obtain
\begin{align}
	P^{(2)}_a = \frac{1}{2} \lim_{\Delta \to \epsilon} \int dz_1 dz_2 \: \mathscr{F}_{\mu_1}(z_1, x+c_1+c_2) \mathscr{F}_{\mu_2}(z_2,x+c_2) A_1(z_1) A_2(z_2).
\end{align}
Here, $c_1 \in \left( \epsilon, \Delta\right)$ and $c_2 \in \left( -\frac{\Delta}{2}, \frac{\Delta}{2} - \epsilon \right)$. Before proceeding with the computation of $P^{(2)}_a$, we verify that the line $z_1 = z_2$ is indeed excluded from the integration domain. To reach this conclusion, we first note that we can choose the support of the $\mathscr{F}$-functions to be of the form:
\begin{align}\label{DM:F_support_condition}
\supp \mathscr{F}_{\mu} (a,b) = \left\{ a,b \in \mathbb{R} \: | \: |a-b| \leq \mu \right\},
\end{align}
in which $\mu \in \mathbb{R}$ act as a regularization parameter. Hence,
\begin{align}
	x + c_1 +c_2 - \mu_1 \leq z_1 \leq x + c_1 + c_2 + \mu_1 \quad \text{and} \quad x + c_2 - \mu_2 \leq z_2 \leq x + c_2 +\mu_2,
\end{align}
so that the condition for disjoint supports corresponds to:
\begin{align}
	x + c_2 + \mu_2 < x+ c_1 + c_2 \quad \Leftrightarrow \quad \mu_1 + \mu_2 < c_1.
\end{align}
Finally, since $c_1 < \epsilon$, it is sufficient to take $\mu_1, \mu_2 < \frac{\epsilon}{2}$ to completely remove the line $z_1=z_2$ from the integration domain.

We can now safely take the limit $\Delta \to \epsilon$,
\begin{align}
	P^{(2)}_a &= \frac{1}{2} \int dz_1 dz_2 \: \mathscr{F}_{\mu_1} \left( z_1, x +\frac{\epsilon}{2} \right) \mathscr{F}_{\mu_2} \left( z_2, x -\frac{\epsilon}{2} \right) A_1(z_1) A_2(z_2) \\
	&= \frac{1}{2} {A_{\scriptscriptstyle\mathscr{F}}}_1 \left( x +\frac{\epsilon}{2} \right) {A_{\scriptscriptstyle\mathscr{F}}}_2 \left( x -\frac{\epsilon}{2} \right). \nonumber
\end{align}
Noting that the term $P^{(2)}_b$ can be similarly evaluated, we conclude that the $\circ$-product of two operator valued distributions is:
\begin{align}\label{DM:2f_product_result}
	P^{(2)} = \frac{1}{2} \left[ {A_{\scriptscriptstyle\mathscr{F}}}_1 \left( x +\frac{\epsilon}{2} \right) {A_{\scriptscriptstyle\mathscr{F}}}_2 \left( x -\frac{\epsilon}{2} \right) + {A_{\scriptscriptstyle\mathscr{F}}}_1 \left( x -\frac{\epsilon}{2} \right) {A_{\scriptscriptstyle\mathscr{F}}}_2 \left( x +\frac{\epsilon}{2} \right)  \right].
\end{align}

Next, we consider the $\circ$-product of four operator valued distributions. This corresponds to:
\begin{align}\label{DM:4f_product}
	P^{(4)} &= {A_{\scriptscriptstyle\mathscr{F}}}_1(x) \circ {A_{\scriptscriptstyle\mathscr{F}}}_2(x) \circ  {A_{\scriptscriptstyle\mathscr{F}}}_3(x) \circ  {A_{\scriptscriptstyle\mathscr{F}}}_4(x) \\
	&=  \int dz_1 dz_2 dz_3 dz_4 \: \fint_x d\xi_1 d\xi_2 d\xi_3 d\xi_4 \: \prod_{j=1}^4F_{\mu_j}(z_j,\xi_j) A_j(z_j)  \nonumber \\
	&= \lim_{\Delta \to 3 \epsilon} \int \frac{dz_1 dz_2 dz_3 dz_4}{(\Delta - 3 \epsilon)^4} \: \sum_{\perm \{\xi\}} \left\{ \int^{x+\frac{\Delta}{2} - 3 \epsilon}_{x - \frac{\Delta}{2}} d\xi_4 \int^{x+ \frac{\Delta}{2} -2 \epsilon}_{\xi_4 + \epsilon} d\xi_3 \int^{x+\frac{\Delta}{2} - \epsilon}_{\xi_3 + \epsilon} d\xi_2 \int^{x+\frac{\Delta}{2}}_{\xi_2 + \epsilon} d\xi_1 \right\} \nonumber \\
	&\cdot \mathscr{F}_{\mu_1} (z_1,\xi_1) \mathscr{F}_{\mu_2} (z_2,\xi_2) \mathscr{F}_{\mu_3} (z_3,\xi_3) \mathscr{F}_{\mu_4} (z_4,\xi_4) A_1(z_1) A_2(z_2) A_3(z_3) A_4(z_4). \nonumber 
\end{align}
Here the integration is taken over a $4$-cube of side $\Delta$ minus a strip of width $2 \epsilon$ around the hyperplanes defined by $\xi_1 = \xi_2$, $\xi_1 = \xi_3$ , $\xi_1 = \xi_4$, $\xi_2 = \xi_3$, $\xi_2 = \xi_4$ and $\xi_3 = \xi_4$. Thus the integration domain corresponds to the union of $4!$ disconnected regions indexed by one of the $4!$ permutations of the variables $\xi_i$, $1\leq i \leq 4$, each of which with a volume of $\Delta V_j = \frac{1}{4!} \left( \Delta - 3 \epsilon\right)^2$, $1 \leq j \leq 4!$. Hence, the limit $\Delta V_j \to 0$ is equivalent to taking $\Delta \to 3 \epsilon$.  

Similarly, by invoking the mean value theorem to compute the integrations over $\xi_1$, $\xi_2$, $\xi_3$ and $\xi_4$, we obtain: 
\begin{align}
	\frac{1}{4!} \lim_{\Delta \to 3 \epsilon}	\int dz_1 dz_2 dz_3 dz_4 \: &\mathscr{F}_{\mu_1}(z_1, x + c_1 + c_2 + c_3 + c_4) \mathscr{F}_{\mu_2}(z_2, x + c_2 + c_3 + c_4) \\
	\cdot& \mathscr{F}_{\mu_3}(z_3, x + c_3 + c_4) \mathscr{F}_{\mu_4}(z_4, x + c_4) A_1(z_1) A_2(z_2) A_3(z_3) A_4(z_4)  \nonumber
\end{align}
for a generic term of the sum over all permutations in $I^{(4)}$. Here, $c_1, c_2, c_3 \in (\epsilon, \Delta - 2 \epsilon)$ and $c_4 \in \left( - \frac{\Delta}{2},\frac{\Delta}{2} - 3 \epsilon \right)$. The conditions for disjoint supports 
\begin{align*}
	x + c_2 + c_3 + c_4 + \mu_2 < x + c_1 + c_2 + c_3 + c_4 - \mu_1 \quad &\Leftrightarrow \quad \mu_1 + \mu_2 < c_1 \\
	x + c_3 + c_4 + \mu_3 < x + c_2 + c_3 + c_4 - \mu_2 \quad &\Leftrightarrow \quad \mu_2 + \mu_3 < c_2 \\
	x + c_4 + \mu_4 < x + c_3 + c_4 - \mu_3 \quad &\Leftrightarrow  \quad \mu_3 + \mu_4< c_3 
\end{align*}
are satisfied whenever $\mu_1, \mu_2, \mu_3, \mu_4 < \frac{\epsilon}{2}$. Therefore, the singular hyperplanes $z_i = z_j$, $i\neq j$, $i,j \in \{ 1,2,3,4\}$ are excluded from the integration domain.

We can now take the limit $\Delta \to 3 \epsilon$ to obtain:
\begin{align}
	\frac{1}{4!} \int dz_1 dz_2 dz_3 dz_4 \: &\mathscr{F}_{\mu_1} \left( z_1, x+ \frac{3 \epsilon}{2} \right) \mathscr{F}_{\mu_2} \left( z_2, x+ \frac{ \epsilon}{2} \right) \mathscr{F}_{\mu_3} \left( z_3, x- \frac{ \epsilon}{2} \right) \\
	\cdot& \mathscr{F}_{\mu_4} \left( z_4, x- \frac{3 \epsilon}{2} \right)  A_1(z_1) A_2(z_2) A_3(z_3) A_4(z_4). \nonumber
\end{align}
Finally, collecting the terms corresponding to all permutations over the arguments of the $\mathscr{F}$-regularized fields, we obtain:
\begin{align}\label{Areg:4f_result}
	P^{(4)} = \frac{1}{4!} \sum_{\perm} {A_{\scriptscriptstyle\mathscr{F}}}_1 \left( x +\frac{3\epsilon}{2} \right) {A_{\scriptscriptstyle\mathscr{F}}}_2 \left( x +\frac{\epsilon}{2} \right) {A_{\scriptscriptstyle\mathscr{F}}}_3 \left( x -\frac{\epsilon}{2} \right) {A_{\scriptscriptstyle\mathscr{F}}}_4 \left( x -\frac{3\epsilon}{2} \right).
\end{align}

\subsection{Quantum Hamiltonian}
From the expression \eqref{DM:sklyanin_prod_general} and the examples involving two and four regularized fields, it is clear that the $\circ$-product  essentially ``smears'' the product of operators around a volume of side $\Delta$ avoiding the singularities at coinciding points. Thus, by combining the field regularization with the operator product regularization, we can formulate a quantum theory with a well-defined algebra of operators and construct the corresponding Hilbert space by diagonalizing the quantum Hamiltonian. Following this prescription, we propose 
\begin{align}\label{DM:quantum_hamiltonian}
	H = \int dx \: \mathscr{H}_{\scriptscriptstyle\mathscr{F}}(x),
\end{align}
with
\begin{align}\label{DM_quantum_hamiltonian_density}
	\mathscr{H}_{\scriptscriptstyle\mathscr{F}}(x) &= i \sigma_z^{jk} {\psi^{\dagger}_{\scriptscriptstyle\mathscr{F}}}_j^{\alpha} (x) \circ \partial_x {\psi_{\scriptscriptstyle\mathscr{F}}}_k^{\alpha} (x) + m \sigma_x^{jk} {\psi^{\dagger}_{\scriptscriptstyle\mathscr{F}}}_j^{\alpha} (x) \circ {\psi_{\scriptscriptstyle\mathscr{F}}}_k^{\alpha} (x) \\
	&- \frac{1}{2} \Lambda^{jl,km} \: G_{\alpha \beta, {\alpha'} {\beta'}} \:{\psi^{\dagger}_{\scriptscriptstyle\mathscr{F}}}_j^{\alpha} (x) \circ {\psi^{\dagger}_{\scriptscriptstyle\mathscr{F}}}_l^{\beta} (x) \circ {\psi_{\scriptscriptstyle\mathscr{F}}}_k^{{\alpha'}} (x) \circ {\psi_{\scriptscriptstyle\mathscr{F}}}_m^{{\beta'}} (x) \nonumber
\end{align}
as the quantum Hamiltonian corresponding to the anisotropic Belavin model \eqref{DM_Original_Hamiltonian}.

Next, we will construct the Hilbert space of the theory by considering the action of  \eqref{DM:quantum_hamiltonian} on the $N$-particle states defined by:
\begin{align}\label{DM:N_particle_state}
	|\psi\rangle_N = \int dx_1 \cdots dx_N \: \phi_{\alpha_1 \cdots \alpha_N}^{i_1 \cdots i_N}(x_1,\ldots, x_N) \: {\psi^{\dagger}_{\scriptscriptstyle\mathscr{F}}}_{i_1}^{\alpha_1} (x_1) \cdots {\psi^{\dagger}_{\scriptscriptstyle\mathscr{F}}}_{i_N}^{\alpha_N} (x_N)  |0\rangle,
\end{align}
with an anti-symmetric wave-function:
\begin{align}\label{DM:antisymmetry_condition}
	\phi^{i_{q_1} \cdots i_{q_N}}_{\alpha_{q_1}\cdots \alpha_{q_N}} (x_{q_1}, \ldots, x_{q_N}) = (-1)^{\sign(q) } \phi_{\alpha_1 \cdots \alpha_N}^{i_1 \cdots i_N}(x_1,\ldots, x_N).
\end{align}
Here, $|0\rangle$ denotes the pseudovacuum, \emph{i.e.}, the state annihilated by the field operator
\begin{align}
	\psi_j^{\alpha} (x) |0\rangle = 0.
\end{align}
The vacuum energy is, therefore, given by $H|0\rangle = 0$. 

Before proceeding with the diagonalization of the 1- and 2-particle sectors, it is important to emphasize that the wave-function $ \phi_{\alpha_1 \cdots \alpha_N}^{i_1 \cdots i_N}(x_1,\ldots, x_N)$ is in general discontinuous and, therefore, not defined on the diagonal $x_{1}=x_{2}=\cdots=x_{N}$. Hence, the integral in \eqref{DM:N_particle_state} should be understood in the sense of removing these points of measure zero. Thus, for example, for the case $N=2$, the integral in \eqref{DM:N_particle_state} should be understood as follows:
\begin{align}\label{DM:eta_cut_off}
	\int dx_1 dx_2 = \int_{x_1 > x_2 } dx_1 dx_2 + \int_{x_1 < x_2 } dx_1 dx_2 = \lim_{\eta \to 0} \left[ \int dx_2 \int_{x_2 + \eta}^{\infty} dx_1+ \int dx_2 \int^{x_2 - \eta}_{-\infty} dx_1 \right].
\end{align}
This leaves open the question of whether one needs to add a compensating ``localized'' state at $x_{1}=x_{2}$. This is indeed the case for the $AAF$ model, as explained in \cite{Melikyan:2014mfa}. Such localized state was shown to be necessary in order to diagonalize the Hamiltonian of the $AAF$ model, and was due to the fact that the corresponding  wave-function was discontinuous together with  its derivatives. Here, however, there is no need to add such a localized term.

\subsubsection{1-Particle Sector}
To evaluate the action of the regularized Hamiltonian \eqref{DM:quantum_hamiltonian} on the 1-particle state
\begin{align}\label{DM:1p-sector}
	H | \psi \rangle_1 =& \int dx dy \: \phi_{\gamma}^i(x) \left[ i \sigma_z^{jk} {\psi^{\dagger}_{\scriptscriptstyle\mathscr{F}}}^{\alpha}_j (y) \circ \partial_y {\psi_{\scriptscriptstyle\mathscr{F}}}^{\alpha}_k(y) + m \sigma_x^{jk} {\psi^{\dagger}_{\scriptscriptstyle\mathscr{F}}}^{\alpha}_j(y) \circ {\psi_{\scriptscriptstyle\mathscr{F}}}^{\alpha}_k(y) \right. \\
	&- \left. \frac{1}{2} \Lambda^{jl,km} G_{\alpha \beta, \tilde{\alpha}\tilde{\beta}} {\psi^{\dagger}_{\scriptscriptstyle\mathscr{F}}}^{\alpha}_j(y) \circ {\psi^{\dagger}_{\scriptscriptstyle\mathscr{F}}}^{\beta}_l(y) \circ {\psi_{\scriptscriptstyle\mathscr{F}}}^{\tilde{\alpha}}_k(y) \circ {\psi_{\scriptscriptstyle\mathscr{F}}}^{\tilde{\beta}}_m(y)\right] {\psi^{\dagger}_{\scriptscriptstyle\mathscr{F}}}^{\gamma}_i (x) |0\rangle, \nonumber
\end{align}
we use the regularized algebra
\begin{align}\label{DM:F_reg_CAR}
	\left\{ {\psi_{\scriptscriptstyle\mathscr{F}}}^{\alpha}_j(x), {\psi^{\dagger}_{\scriptscriptstyle\mathscr{F}}}^{\beta}_k(y) \right\} = \delta^{\alpha \beta} \delta{jk} \int d \xi \: \mathscr{F}_{\mu} (\xi,x) \mathscr{F}_{\nu}(\xi,y)
\end{align}
to normal order each term in \eqref{DM:1p-sector}. 

We start with the kinetic term
\begin{align}
	K^{(1)} &= \int dx dy \: \phi^i_{\gamma}(x) \:{\psi^{\dagger}_{\scriptscriptstyle\mathscr{F}}}^{\alpha}_j (y) \circ \partial_y {\psi_{\scriptscriptstyle\mathscr{F}}}^{\alpha}_k(y) \: {\psi^{\dagger}_{\scriptscriptstyle\mathscr{F}}}^{\gamma}_i (x) |0\rangle \\
	&= \delta^{\alpha \gamma} \delta_{ij} \int dx dy d\xi \fint_y du dv \: \phi^i_{\gamma}(x) \: \partial_v \mathscr{F}_{\mu} (v, \xi) \mathscr{F}_{\nu} (x,\xi) \: {\psi^{\dagger}_{\scriptscriptstyle\mathscr{F}}}^{\alpha}_j (u)  |0\rangle. \nonumber
\end{align}
In order to reduce it to the form \eqref{DM:2f_product}, we use the anti-symmetry of the derivatives of the $\mathscr{F}$-functions
\begin{align}\label{DM:F_function_derivative_property}
	\partial_v \mathscr{F}_{\mu} (v,u) = -\partial_u \mathscr{F}_{\mu} (v,u)
\end{align}
together with their rapidly decreasing behavior to move the derivative from the $\mathscr{F}$-functions to the wave-function through a series of integrations by parts:
\begin{align}
	 K^{(1)} = \delta^{\alpha \gamma} \delta_{ij} \int dx dy d\xi \fint_y du dv \: \partial_x\phi^i_{\gamma}(x) \:  \mathscr{F}_{\mu} (v, \xi) \mathscr{F}_{\nu} (x,\xi) \: {\psi^{\dagger}_{\scriptscriptstyle\mathscr{F}}}^{\alpha}_j (x)  |0\rangle.
\end{align}
Thus, we can integrate over $u$ and $v$ following the steps outlined in the section \ref{Quantization} arriving at:
\begin{align}
	K^{(1)} = \frac{1}{2} \delta^{\alpha \gamma} \delta_{ij} \int dx dy d\xi &\left[ \partial_x \phi_{\gamma}^i(x) \mathscr{F}_{\mu} \left( y -\frac{\epsilon}{2}, \xi \right) \mathscr{F}_{\nu}(x, \xi) {\psi^{\dagger}_{\scriptscriptstyle\mathscr{F}}}^{\alpha}_j\left( y +\frac{\epsilon}{2} \right) \right. \\
	+& \left. \partial_x \phi_{\gamma}^i(x) \mathscr{F}_{\mu} \left( y +\frac{\epsilon}{2}, \xi \right) \mathscr{F}_{\nu}(x, \xi) {\psi^{\dagger}_{\scriptscriptstyle\mathscr{F}}}^{\alpha}_j\left( y -\frac{\epsilon}{2} \right) \right] |0\rangle \nonumber.
\end{align}
There remains to remove the regularization parameters $\mu$ and $\nu$. Since $\mu, \nu < \frac{\epsilon}{2}$, we can simply take the limits $\mu, \nu \to 0$ using \eqref{DM:F_limit} to obtain:
\begin{align}
	K^{(1)} = \zeta^2 \: \delta^{\alpha \gamma} \delta_{ij} \int dx \: \partial_x \phi_{\gamma}^i (x) {\psi^{\dagger}_{\scriptscriptstyle\mathscr{F}}}^{\alpha}_j(x) |0\rangle.
\end{align}

The mass term in \eqref{DM:1p-sector} is already in the form \eqref{DM:2f_product}, so we can simply use the formula \eqref{DM:2f_product_result} to get:
\begin{align}
	M^{(1)} =& \int dx dy \: \phi_{\gamma}^i(x) \: {\psi^{\dagger}_{\scriptscriptstyle\mathscr{F}}}^{\alpha}_j(y) \circ {\psi_{\scriptscriptstyle\mathscr{F}}}^{\alpha}_k(y) \: {\psi^{\dagger}_{\scriptscriptstyle\mathscr{F}}}^{\gamma}_i (x) |0\rangle \\
	=& \frac{1}{2} \delta^{\alpha \gamma} \delta_{ij} \int dx dy d\xi \: \phi_{\gamma}^i(x) \mathscr{F}_{\nu} (x, \xi)) \left[ \mathscr{F}_{\mu} \left( y + \frac{\epsilon}{2}, \xi \right) {\psi^{\dagger}_{\scriptscriptstyle\mathscr{F}}}^{\alpha}_j \left( y - \frac{\epsilon}{2} \right) \right. \nonumber \\
	&+ \left.  \mathscr{F}_{\mu} \left( y - \frac{\epsilon}{2}, \xi \right) {\psi^{\dagger}_{\scriptscriptstyle\mathscr{F}}}^{\alpha}_j \left( y + \frac{\epsilon}{2} \right) \right|0\rangle. \nonumber
\end{align}
Then, removing the regularization parameters $\mu,\nu$, we can perform the integration over $y$ and $\xi$:
\begin{align}
	M^{(1)} = \zeta^2 \:  \delta^{\alpha \gamma} \delta_{ij} \int dx \: \phi_{\gamma}^{i}(x) {\psi^{\dagger}_{\scriptscriptstyle\mathscr{F}}}^{\alpha}_j (x) |0\rangle.
\end{align}

Finally, noting that the interaction term in  \eqref{DM:1p-sector} is of the fourth order in the fields, we can easily conclude from the algebra \eqref{DM:F_reg_CAR} that it trivially vanishes. Hence, we are left we the following expression for the action of the hamiltonian \eqref{DM:quantum_hamiltonian} on the $1$-particle state:
\begin{align}\label{DM:1p_action_f}
	H | \psi\rangle_1 = \zeta^2 \int dx \: \left[ i \sigma_z^{jk} \partial_x + m \sigma_x^{jk} \right] \phi^k_{\alpha} (x) \; {\psi^{\dagger}_{\scriptscriptstyle\mathscr{F}}}^{\alpha}_j (x) |0\rangle.
\end{align}

The wave functions describing the 1-particle state \eqref{DM:N_particle_state} can be derived by solving the differential equation:
\begin{align}
	H | \psi\rangle_1 = E_1 |\psi\rangle_1 \quad \Leftrightarrow \quad \left[ i \sigma_z^{jk} \partial_x + m \sigma_x^{jk} \right] \phi^k_{\alpha} (x) = E_1 \phi^k_{\alpha},
\end{align}
following from \eqref{DM:1p_action_f}. The solution corresponding to the positive mass shell,
\begin{align}\label{DM:1p_positive_mass_shell}
	\phi_{\alpha}^j(x) = A_{\alpha} u^j(\theta) e^{i k x}, \quad \text{with} \quad E_1 = m \cosh \theta, \; k = m \sinh \theta,
\end{align}
is parametrized by a real rapidity $\theta$. Here, $k$ denotes the wave vector and $u (\theta)$ is the Dirac spinor:
\begin{align}\label{DM:1p_dirac_spinor}
	u(\theta) = \frac{1}{\sqrt{2 \cosh \theta}} \begin{pmatrix}e^{-\frac{\theta}{2}} \\ e^{\frac{\theta}{2}} \end{pmatrix}.
\end{align}
On the other hand, the negative mass shell can be trivially obtained from \eqref{DM:1p_dirac_spinor} after performing the transformation: $\theta \to i \pi - \theta$.

\subsubsection{2-Particle Sector}
Next, we consider the action of the regularized Hamiltonian \eqref{DM:quantum_hamiltonian} on the 2-particle state. Using the property \eqref{DM:F_function_derivative_property} and integrating by parts to remove all derivatives acting on $\mathscr{F}$-functions, we obtain:
\begin{align} \label{DM:2p-sector}
	H | \psi\rangle_2 = \left\{ 2i \delta^{\alpha \alpha_1} \delta_{k i_1} \left[  \sigma_z^{jk}  K^{(2)} + m \sigma_x^{jk} M^{(2)} - \sigma_z^{jk} \left( B^{(2)}_{\xi} + B^{(2)}_{x}\right)  \right] + \Lambda^{jl,i_1i_2} G_{\alpha \beta, \alpha_1 \alpha_2}I^{(2)}  \right\} |0\rangle.
\end{align}
Here,
\begin{align} 
	B^{(2)}_{\xi} &=  \int dy dx_1 dx_2 d \xi \fint_y du_1 du_2 \; \partial_{\xi} \left[ \phi^{i_1i_2}_{\alpha_1 \alpha_2} (x_1,x_2) \mathscr{F}_{\mu}(x_1, \xi)  \mathscr{F}_{\nu} (u_2,\xi) {\psi^{\dagger}_{\scriptscriptstyle\mathscr{F}}}^{\alpha}_j(u_1) {\psi^{\dagger}_{\scriptscriptstyle\mathscr{F}}}^{\alpha_2}_{i_2}(x_2) \right] \label{DM:2p_Bxi},  \\
	B^{(2)}_{x} &=  \int dy dx_1 dx_2 d \xi \fint_y du_1 du_2 \;  \partial_{x_1} \left[ \phi^{i_1i_2}_{\alpha_1 \alpha_2} (x_1,x_2) \mathscr{F}_{\mu}(x_1, \xi)  \mathscr{F}_{\nu} (u_2,\xi) {\psi^{\dagger}_{\scriptscriptstyle\mathscr{F}}}^{\alpha}_j(u_1) {\psi^{\dagger}_{\scriptscriptstyle\mathscr{F}}}^{\alpha_2}_{i_2}(x_2) \right] \label{DM:2p_Bx}
\end{align}
correspond to the boundary terms arising from the integrations by parts, while the kinetic, mass and interaction terms are given respectively by
\begin{align}
	K^{(2)} =& \int dy dx_1 dx_2 d \xi \fint_y du_1 du_2 \; \partial_{x_1} \phi^{i_1i_2}_{\alpha_1 \alpha_2} (x_1,x_2) \mathscr{F}_{\mu}(x_1, \xi)  \mathscr{F}_{\nu} (u_2,\xi) {\psi^{\dagger}_{\scriptscriptstyle\mathscr{F}}}^{\alpha}_j(u_1) {\psi^{\dagger}_{\scriptscriptstyle\mathscr{F}}}^{\alpha_2}_{i_2}(x_2), \label{DM:2p_K}\\
	M^{(2)} =&  \int dy dx_1 dx_2 d \xi \fint_y du_1 du_2 \;  \phi^{i_1i_2}_{\alpha_1 \alpha_2} (x_1,x_2) \mathscr{F}_{\mu}(x_1, \xi)  \mathscr{F}_{\nu} (u_2,\xi) {\psi^{\dagger}_{\scriptscriptstyle\mathscr{F}}}^{\alpha}_j(u_1) {\psi^{\dagger}_{\scriptscriptstyle\mathscr{F}}}^{\alpha_2}_{i_2}(x_2), \label{DM:2p_M}\\
	I^{(2)} =& \int dy dx_1 dx_2 d\xi_1 d\xi_2 \fint_y du_1 du_2 du_3 du_4 \; \phi^{i_1i_2}_{\alpha_1 \alpha_2} (x_1,x_2) \mathscr{F}_{\mu_1} (x_1,\xi_1) \mathscr{F}_{\mu_2} (x_2,\xi_2) \label{DM:2p_I}\\
	&\cdot \mathscr{F}_{\nu_3}(u_3,\xi_1) \mathscr{F}_{\nu_4}(u_4,\xi_2) {\psi^{\dagger}_{\scriptscriptstyle\mathscr{F}}}^{\alpha}_{j}(u_1) {\psi^{\dagger}_{\scriptscriptstyle\mathscr{F}}}^{\beta}_{l}(u_2). \nonumber  
\end{align}
The derivation of \eqref{DM:2p-sector} involves integrating by parts with respect to $x_1$. Since before diagonalizing the Hamiltonian, we cannot know the exact dependence of the wave function or its derivatives on their arguments, we must ensure that all possible singularities arising from the line $x_1=x_2$ be taken into consideration accordingly. Therefore, all integrals with respect to $x_1$ and $x_2$ should be carefully evaluated on the line $x_1 = x_2$. In the following, we will discuss in more details how this can be consistently achieved as we evaluate the contributions from \eqref{DM:2p_Bx} and \eqref{DM:2p_K}.

We start by considering the simpler boundary term $B^{(2)}_{\xi}$. In this case, we can trivially exchange the integration over $u_1$ and $u_2$ with the derivative with respect to $\xi$ to obtain an expression of the form \eqref{DM:2f_product}. So that, we can follow the steps outlined in the section \ref{Quantization} to conclude that this boundary term is proportional to:
\begin{align}
	B^{(2)}_{\xi} &\propto \int d \xi \: \partial_{\xi} \left[ \mathscr{F}_{\mu} (x_1,\xi) \mathscr{F}_{\nu} \left( y \pm \frac{\epsilon}{2}, \xi \right) \right] \\
	&= \lim_{a  \to \infty} \left[  \mathscr{F}_{\mu} (x_1, a) \mathscr{F}_{\nu} \left( y \pm \frac{\epsilon}{2}, a \right) - \mathscr{F}_{\mu} (x_1, -a) \mathscr{F}_{\nu} \left( y \pm \frac{\epsilon}{2}, -a \right) \right] \nonumber.
\end{align}
For fixed $x_1 = z$ or $y = z$ and sufficiently small $\mu$, the support condition of the $\mathscr{F}$-functions \eqref{DM:F_support_condition} implies that $a \notin \supp \mathscr{F}_{\mu} (z,a)$ in the limit $a \to \infty$. Hence, this boundary term must trivially vanish, \emph{i.e.}, 
\begin{align} \label{DM:2p_Bxi_f}
B^{(2)}_{\xi}  = 0.
\end{align}

The second boundary term, $B^{(2)}_{x}$, requires a more careful consideration involving the explicit use of \eqref{DM:eta_cut_off} because of the possible discontinuities of the wave-function and its derivatives. Taking that into account, we can proceed as before and exchange the derivative with respect to $x_1$ with the integration over $u_1$ and $u_2$ to obtain: 
\begin{align}
	B^{(2)}_{x} =  \frac{1}{2} \int dy dx_1 dx_2 d\xi \: \partial_{x_1} &\left\{ \phi^{i_1 i_2}_{\alpha_1 \alpha_2} (x_1,x_2) \mathscr{F}_{\mu} (x_1, \xi) \left[ \mathscr{F}_{\nu}\left( y - \frac{\epsilon}{2}, \xi \right) {\psi^{\dagger}_{\scriptscriptstyle\mathscr{F}}}^{\alpha}_{j}\left( y+\frac{\epsilon}{2} \right) \right. \right. \\
	&+ \left. \left. \mathscr{F}_{\nu}\left( y + \frac{\epsilon}{2}, \xi \right) {\psi^{\dagger}_{\scriptscriptstyle\mathscr{F}}}^{\alpha}_{j}\left( y-\frac{\epsilon}{2} \right)  \right] {\psi^{\dagger}_{\scriptscriptstyle\mathscr{F}}}^{\alpha_2}_{i_2}(x_2)  \right\}. \nonumber
\end{align}
After integrating the total derivative and removing the regularization parameters $\mu, \nu < \frac{\epsilon}{2}$, we obtain:
\begin{align}
	B^{(2)}_{x} =  \frac{\zeta^2}{2}  \int dy &\left[ \phi^{i_1 i_2}_{\alpha_1 \alpha_2} \left( y- \frac{\epsilon}{2}, y- \frac{\epsilon}{2} + \eta \right) {\psi^{\dagger}_{\scriptscriptstyle\mathscr{F}}}^{\alpha}_{j} \left( y + \frac{\epsilon}{2} \right) {\psi^{\dagger}_{\scriptscriptstyle\mathscr{F}}}^{\alpha_2}_{i_2} \left( y -\frac{\epsilon}{2} +\eta \right) \right. \\
	+& \left. \phi^{i_1 i_2}_{\alpha_1 \alpha_2} \left( y+ \frac{\epsilon}{2}, y+ \frac{\epsilon}{2} + \eta \right) {\psi^{\dagger}_{\scriptscriptstyle\mathscr{F}}}^{\alpha}_{j} \left( y - \frac{\epsilon}{2} \right) {\psi^{\dagger}_{\scriptscriptstyle\mathscr{F}}}^{\alpha_2}_{i_2} \left( y +\frac{\epsilon}{2} +\eta \right) \right. \nonumber \\
	-& \left. \phi^{i_1 i_2}_{\alpha_1 \alpha_2} \left( y- \frac{\epsilon}{2}, y- \frac{\epsilon}{2} - \eta \right) {\psi^{\dagger}_{\scriptscriptstyle\mathscr{F}}}^{\alpha}_{j} \left( y + \frac{\epsilon}{2} \right) {\psi^{\dagger}_{\scriptscriptstyle\mathscr{F}}}^{\alpha_2}_{i_2} \left( y -\frac{\epsilon}{2} -\eta \right) \right. \nonumber \\
	-& \left. \phi^{i_1 i_2}_{\alpha_1 \alpha_2} \left( y+ \frac{\epsilon}{2}, y+ \frac{\epsilon}{2} - \eta \right) {\psi^{\dagger}_{\scriptscriptstyle\mathscr{F}}}^{\alpha}_{j} \left( y - \frac{\epsilon}{2} \right) {\psi^{\dagger}_{\scriptscriptstyle\mathscr{F}}}^{\alpha_2}_{i_2} \left( y +\frac{\epsilon}{2} -\eta \right) \right]. \nonumber 
\end{align}
We can now shift the integration over $y$, so that it is possible to use the continuity of the fields on their arguments to remove the cut off $\eta$. The resulting contribution from the boundary terms amount to:
\begin{align}\label{DM:2p_Bx_f}
	B^{(2)}_{x} =  \zeta^2 \int dx \: \left[ \phi^{i_1 i_2}_{\alpha_1 \alpha_2}(x - \epsilon, x) - \phi^{i_1 i_2}_{\alpha_1 \alpha_2}(x + \epsilon, x) \right]  {\psi^{\dagger}_{\scriptscriptstyle\mathscr{F}}}^{\alpha}_j(x)  {\psi^{\dagger}_{\scriptscriptstyle\mathscr{F}}}^{\alpha_2}_{i_2}(x)  .
\end{align}

The kinetic \eqref{DM:2p_K} and mass \eqref{DM:2p_M} terms require a similar careful analysis regarding the integration over $x_1$ and $x_2$. Following exactly the same steps as before, we obtain:
\begin{align}\label{DM:2p_K_f}
	K^{(2)} = \zeta^2 \int_{x_1 \neq x_2} dx_1 dx_2 \; \partial_{x_1} \phi^{i_1 i_2}_{\alpha_1 \alpha_2}(x_1, x_2) \;{\psi^{\dagger}_{\scriptscriptstyle\mathscr{F}}}^{\alpha}_j(x_1)  {\psi^{\dagger}_{\scriptscriptstyle\mathscr{F}}}^{\alpha_2}_{i_2}(x_2), 
\end{align}
for the kinetic term and
\begin{align}\label{DM:2p_M_f}
	M^{(2)} = \zeta^2 \int_{x_1 \neq x_2} dx_1 dx_2 \; \phi^{i_1 i_2}_{\alpha_1 \alpha_2} (x_1, x_2) \;{\psi^{\dagger}_{\scriptscriptstyle\mathscr{F}}}^{\alpha}_j(x_1)  {\psi^{\dagger}_{\scriptscriptstyle\mathscr{F}}}^{\alpha_2}_{i_2}(x_2),  
\end{align}
for the mass term.

Finally, the contribution of the interaction term \eqref{DM:2p_I} can be evaluated by following the prescription outlined in the section \ref{Quantization}. In this case, the interplay between the $\circ$-product and $\mathscr{F}$-regularization guarantees that the singular hyperplane $x_1 = x_2$ is never reached without any further consideration. More precisely, the support condition of the $\mathscr{F}$-functions \eqref{DM:F_support_condition} together with the requirement that $\mu_1, \mu_2, \nu_3, \nu_4 < \frac{\epsilon}{2}$ naturally exclude this hyperplane from the domain of integration. Thus,
\begin{align}\label{DM:2p_I_perm}
	I^{(2)} = \frac{\zeta^4}{24} \sum_{\text{perm}} \int dx \; \phi^{i_1 i_2}_{\alpha_1 \alpha_2} \left( u_3, u_4  \right) {\psi^{\dagger}_{\scriptscriptstyle\mathscr{F}}}^{\alpha}_j \left( u_1\right) {\psi^{\dagger}_{\scriptscriptstyle\mathscr{F}}}^{\beta}_l \left( u_2\right),
\end{align}
where the permutation is taken over $u_1$, $u_2$, $u_3$ and $u_4$ which depend both on $x$ and on the regularizing parameter $\epsilon$ as:
\begin{align}
	u_1 = x + \frac{3 \epsilon}{2}, \;  u_2 = x + \frac{\epsilon}{2}, \; u_3 = x - \frac{\epsilon}{2}, \; u_4 = x - \frac{3 \epsilon}{2}.
\end{align}
To simplify the sum over all permutations appearing in \eqref{DM:2p_I_perm}, we shifted the integration variables so that all terms had wave-functions with symmetrical arguments. Then, we rescaled the regularizing parameter for each term individually and used the continuity of the fields on their arguments to partially evaluate the limit $\epsilon \to 0$, leading to:
\begin{align}\label{DM:2p_I_f}
	I^{(2)} = \frac{\zeta^4}{2} \int dx \; \left[  \phi^{i_1 i_2}_{\alpha_1 \alpha_2} \left( x - \epsilon, x \right) +  \phi^{i_1 i_2}_{\alpha_1 \alpha_2} \left( x + \epsilon, x \right) \right] {\psi^{\dagger}_{\scriptscriptstyle\mathscr{F}}}^{\alpha}_j \left( x \right) {\psi^{\dagger}_{\scriptscriptstyle\mathscr{F}}}^{\beta}_l \left( x \right).
\end{align}

Substituting the expressions \eqref{DM:2p_Bxi_f}, \eqref{DM:2p_Bx_f}, \eqref{DM:2p_K_f}, \eqref{DM:2p_K_f} and \eqref{DM:2p_I_f} back into \eqref{DM:2p-sector}, we obtain the action of the quantum Hamiltonian \eqref{DM:quantum_hamiltonian} on the two-particle state:
\begin{align}\label{DM:2p_action_f}
	H | \psi \rangle_2 &= 2 \zeta^2 \int_{x_1 \neq x_2} dx_1 dx_2 \left[ i \sigma_z^{ji_1} \partial_{x_1} + m \sigma_x^{ji_1}  \right]  \phi^{i_1i_2}_{\alpha \beta} (x_1, x_2) {\psi^{\dagger}_{\scriptscriptstyle\mathscr{F}}}^{\alpha}_j(x_1)  {\psi^{\dagger}_{\scriptscriptstyle\mathscr{F}}}^{\beta}_{i_2}(x_2) |0\rangle \\
	&+ \int dx \left[ 2i \zeta^2 \sigma_z^{jk}  {\Delta^{(-)}}_{\alpha \beta}^{kl} + \frac{\zeta^4}{2} \Lambda^{jl,i_1i_2} \; G_{\alpha \beta, \alpha_1 \alpha_2}  {\Delta^{(+)}}_{\alpha_1 \alpha_2}^{i_1 i_2} \right]  {\psi^{\dagger}_{\scriptscriptstyle\mathscr{F}}}^{\alpha}_j(x)  {\psi^{\dagger}_{\scriptscriptstyle\mathscr{F}}}^{\beta}_{l}(x) |0\rangle. \nonumber
\end{align}
Here, we introduced the following shorthand notation for the relevant combination of wave-functions with respect to the dependence on the regularizing parameter:
\begin{align}\label{DM:2p_Delta}
	{\Delta^{(\pm)}}^{ij}_{\alpha \beta} = \phi^{ij}_{\alpha \beta}(x + \epsilon,x) \pm \phi^{ij}_{\alpha \beta}(x - \epsilon,x).
\end{align}
It follows from the anti-symmetry of the wave function and the continuity of the fields that the afore defined quantity ${\Delta^{(\pm)}}^{ij}_{\alpha \beta}$ satisfies the following property:
\begin{align}\label{DM:2p_Delta_symmetry}
	{\Delta^{(\pm)}}^{ij}_{\alpha \beta} = \mp {\Delta^{(\pm)}}^{ji}_{\beta \alpha}. 
\end{align}

In the remainder of this section, we consider the necessary conditions for diagonalizing the 2-particle sector. From the vanishing of the second line in \eqref{DM:2p_action_f}, we obtain the following equations describing the discontinuities of the 2-particle wave function:
\begin{align} 
	i {\Delta^{(-)}}^{12}_{11} + 2 \tilde{\xi} (g_0 + g_3) {\Delta^{(+)}}^{12}_{11} + 2 \tilde{\xi} (g_1 - g_2) {\Delta^{(+)}}^{12}_{22} &=0, \label{DM:2p_syseq1}\\
	i {\Delta^{(-)}}^{12}_{12} + 2 \tilde{\xi} (g_0 - g_3) {\Delta^{(+)}}^{12}_{12} + 2 \tilde{\xi} (g_1 + g_2) {\Delta^{(+)}}^{12}_{21} &=0, \label{DM:2p_syseq2} \\
	i {\Delta^{(-)}}^{12}_{21} + 2 \tilde{\xi} (g_0 - g_3) {\Delta^{(+)}}^{12}_{21} + 2 \tilde{\xi} (g_1 + g_2) {\Delta^{(+)}}^{12}_{12} &=0,\label{DM:2p_syseq3}\\
	i {\Delta^{(-)}}^{12}_{22} + 2 \tilde{\xi} (g_0 + g_3) {\Delta^{(+)}}^{12}_{22} + 2 \tilde{\xi} (g_1 - g_2) {\Delta^{(+)}}^{12}_{11} &=0, \label{DM:2p_syseq4}
\end{align}
where for further convenience we introduced $\tilde{\xi} = \frac{\zeta^2}{4}$. All the remaining combinations ${\Delta^{(\pm)}}^{ij}_{\alpha \beta}$ are continuous by virtue of \eqref{DM:2p_Delta_symmetry}.

To solve the system of equations \eqref{DM:2p_syseq1} -- \eqref{DM:2p_syseq4}, we invoke the Bethe Ansatz to write the 2-particle wave functions as:
\begin{align}\label{DM:2p_BA1}
	{\phi^{(12)}}^{ij}_{\alpha_1 \alpha_2} (x_1,x_2) = A^{12}_{\alpha_1 \alpha_2} u^i \left(\theta_1\right) u^j \left( \theta_2 \right) e^{i\left( k_1 x_1 + k_2 x_2 \right)} - A^{21}_{\alpha_1 \alpha_2} u^i \left(\theta_2\right) u^j \left( \theta_1 \right) e^{i\left( k_2 x_1 + k_1 x_2 \right)}, 
\end{align}	
for $x_1 < x_2$, and
\begin{align}\label{DM:2p_BA2}
	{\phi^{(21)}}^{ij}_{\alpha_1 \alpha_2} (x_1,x_2) = A^{21}_{\alpha_2 \alpha_1} u^i \left(\theta_1\right) u^j \left( \theta_2 \right) e^{i\left( k_1 x_1 + k_2 x_2 \right)} - A^{12}_{\alpha_2 \alpha_1} u^i \left(\theta_2\right) u^j \left( \theta_1 \right) e^{i\left( k_2 x_1 + k_1 x_2 \right)}, 
\end{align}
for $x_1 > x_2$. Here $k_1$ and $k_2$ are wave vectors, $A^{ij}_{\alpha \beta}$ are the unknown $A$-amplitudes and $u(\theta)$ is the Dirac spinor \eqref{DM:1p_dirac_spinor}. Substituting this Ansatz in the system \eqref{DM:2p_syseq1} -- \eqref{DM:2p_syseq4}, we obtain the following system of equations:
\begin{align}
	i \tanh \theta_{12} \left( B_0^{12} + B_0^{21} \right) + \lambda_0 \left( B_0^{12} - B_0^{21} \right) &= 0, \label{DM:2p_syseqB1}\\
	i \coth \theta_{12} \left( B_1^{21} - B_1^{12} \right) - \lambda_1 \left( B_1^{21} + B_1^{12} \right) &= 0, \label{DM:2p_syseqB2}\\
	i \coth \theta_{12} \left( B_2^{21} - B_2^{12} \right) - \lambda_2 \left( B_2^{21} + B_2^{12} \right) &=0, \label{DM:2p_syseqB3}\\ 
	i \coth \theta_{12} \left( B_3^{21} - B_3^{12} \right) - \lambda_3 \left( B_3^{21} + B_3^{12} \right) &= 0, \label{DM:2p_syseqB4}
\end{align}
where $\theta_{ij} = \frac{1}{2} \left( \theta_i - \theta_j \right)$, for the $B$-amplitudes defined as:
\begin{align}\label{DM:2p_B-amplitudes}
	B_0^{ij} = A_{12}^{ij} - A_{21}^{ij}, \quad B_2^{ij} = A_{11}^{ij} + A_{22}^{ij}, \quad B_3^{ij} = A_{11}^{ij} - A_{22}^{ij}, \quad \text{and} \quad B_4^{ij} = A_{12}^{ij} + A_{21}^{ij}.
\end{align} 
For the sake of clarity, we introduced the following combinations of the coupling constants:
\begin{align}
	\lambda_0 &= 2 \tilde{\xi} \left( g_0 - g_1 - g_2 - g_3 \right), \label{DM:2p_lambda1}\\
	\lambda_1 &= 2 \tilde{\xi} \left( g_0 + g_1 - g_2 + g_3 \right), \label{DM:2p_lambda2} \\
	\lambda_2 &= 2 \tilde{\xi} \left( g_0 - g_1 + g_2 + g_3 \right), \label{DM:2p_lambda3} \\
	\lambda_3 &= 2 \tilde{\xi} \left( g_0 + g_1 + g_2 - g_3 \right). \label{DM:2p_lambda4}
\end{align}
Note that the $\lambda_{\mu}$ constants will enter into expressions for physicial amplitudes, and in order to obtain finite expressions for the $S$-matrices one has  to appropriately renormalize the ``bare" coupling constants $g_{0},\ldots,g_{3}$. This will be considered in the subsequent section.  

Thus, the $B$-amplitudes are related as
\begin{align} \label{DM:2p_B_amp_general_relations}
	B_{\mu}^{12} = \gamma_{\mu} \left(\theta_{12} \right) B_{\mu}^{21},
\end{align}
with the $\gamma_{\mu} (\theta)$, $0 \leq \mu \leq 3$ coefficients given by:
\begin{align}\label{DM:2p_gamma_coefs}
	\gamma_0 (\theta) = - \frac{1 + i \lambda_0 \coth \theta}{1 - i \lambda_0 \coth \theta} \quad \text{and} \quad \gamma_j (\theta) =  \frac{\coth \theta + i \lambda_j}{\coth \theta - i \lambda_j}, \: j=1,2,3.
\end{align}

The $n \geq 3$ case will be considered in section \ref{section:8vertex} in relation to the $S$-matrix factorization property, implying the quantum integrability of the model. First, however, we briefly explain in the next section how to generalize the results from the previous analysis by introducing additional interaction terms, the form of which are stipulated by the corresponding interaction terms of the $AAF$ model.  Although this was our main motivation, one can in principle easily modify the analysis below by considering different, more general interaction terms.

\section{Extended anisotropic interactions}
\label{section:Extensions}
As discussed in the introduction, our main motivation is to find an integrable massive fermionic model which contains no derivatives of the  fields, and which may produce the $AAF$ model in the low-energy limit, possibly after a suitable redefinition of the fields. This stems from the observation that the interaction term of the corresponding Hamiltonian has the form of a decomposition in $\nicefrac{1}{m}$ and goes up to the third order, resembling a low-energy expansion of a more fundamental theory.\footnote{We do not reproduce here the explicit lenghty expression for the Hamiltonian, which can be found in \cite{Melikyan:2016gkd}.} Recall, that the Lagrangian of the $AAF$ model has the form:
\begin{align}
	 \mathscr{L}_{\textrm{AAF}} &= i \bar{\psi} \gamma_{\mu} \partial^{\mu} \psi \: - m \bar{\psi} \psi + \frac{g_2}{4m} \epsilon^{\alpha \beta} \left( \bar{\psi}
	\partial_{\alpha} \psi \; \bar{\psi}\: \gamma^3 
	\partial_{\beta} \psi -
	\partial_{\alpha}\bar{\psi} \psi \; 
	\partial_{\beta} \bar{\psi}\: \gamma^3 \psi \right)- \nonumber \\
	&- \frac{g_3}{16m} \epsilon^{\alpha \beta} \left(\bar{\psi}\psi\right)^2 
	\partial_{\alpha}\bar{\psi}\:\gamma^3
	\partial_{\beta}\psi. \label{ext:aaf_lagrangian}
\end{align}
We note here, that, as was shown in \cite{Melikyan:2011uf}, the $S$-matrix factorization, implying the quantum integrability of the model, occurs only when the condition $(g_{2})^{2}=g_{3}$ is satisfied.

As a first step in this direction, we extend in this section the Lagrangian \eqref{DM_Original_Lagrangian} to include two new interaction terms, with the coupling constants $g_{4}$ and $g_{5}$, that resemble that of the $AAF$ model. Thus, we consider the following model:\footnote{We emphasize that although we use the same notation for the fermionic field in the Lagrangians \eqref{ext:aaf_lagrangian} and \eqref{ext:Lagrangian_g4g5}, $\psi$ in the former Lagrangian is a two-component spinor, while in the latter Lagrangian it has four independent components.}
\begin{align}\label{ext:Lagrangian_g4g5}
	\mathscr{L} &= i \bar{\psi} \gamma_{\mu} \partial^{\mu} \psi - m \bar{\psi} \psi - \frac{g_0}{2} \left( \bar{\psi} \gamma^{\mu} \psi \right)  \left( \bar{\psi} \gamma_{\mu} \psi \right) -\frac{g_a}{2}  \left( \bar{\psi} \gamma^{\mu} \tau^a \psi \right) \left( \bar{\psi} \gamma_{\mu} \tau^a \psi \right)\notag \\
	&-\frac{g_{4}}{2}\left( \bar{\psi} \gamma^{\mu} \psi \right) \left( \bar{\psi} \gamma_{\mu} \tau^3 \psi \right)
	-\frac{g_{5}}{2}\varepsilon_{\alpha \beta}\left( \bar{\psi} \gamma^{\alpha} \psi \right) \left( \bar{\psi} \gamma^{\beta} \tau^3 \psi \right).
\end{align}
Note that, without considering anisotropic extensions of the Thirring model, it would have been impossible to write a term proportional to $\varepsilon_{\alpha \beta}$. We stress that we do not try here to find a Lagrangian that would exactly reproduce the $AAF$ Lagrangian \eqref{ext:aaf_lagrangian}, but merely to consider the consequences on the intregrability by adding some terms that resemble the interaction terms in \eqref{ext:aaf_lagrangian}. Namely, in this case, the two new terms in \eqref{ext:Lagrangian_g4g5}, anisotropic with respect to $\tau^{3}$ matrix, were added to resemble those quartic in the fermionic field interaction terms of the $AAF$ Lagrangian \eqref{ext:aaf_lagrangian}. These terms indeed satisfy our requirements of absence of the derivatives in the interaction terms. The natural question is then how such derivative terms may be obtained from \eqref{ext:Lagrangian_g4g5}, and reproduce those of \eqref{ext:aaf_lagrangian}. First, it is easy to see that such derivative terms will appear upon choosing two fields to be integrated out. Another possible way to relate the two Lagrangians is to consider a field transformation with the generic form: $\psi \longrightarrow \psi + \gamma^{\alpha}\partial_{\alpha}\psi + \cdots$. It is clear that amongst the resulting terms after such transformation one will find exactly a term corresponding to the quartic terms in \eqref{ext:aaf_lagrangian}. Using the so-called equivalence theorem (see \cite{Melikyan:2014mfa} for details), one can show that the $S$-matrix is unchanged under such field transformations, and moreover, the quartic terms are essentially the ones determining the complete $S$-matrix for an integrable system. These general arguments will be considered in details elsewhere.

The analysis of the previous section concerning the 1- and 2-particle sectors of the anisotropic Belavin model can be easily generalized. The original tensors $G_{\alpha \beta, {\alpha'} {\beta'}}$ and $ \Lambda^{jl,km}$ (c.f. \eqref{DM:G_tensor_orig} and \eqref{DM:Lambda_tensor_orig}) should now be split into two pairs to accommodate the new interaction terms, the first, corresponding to terms proportional to $g_{0},\ldots, g_{4}$, is
\begin{align}
	G_{\alpha \beta, {\alpha '} {\beta '}} &= \left( g_0 \mathbb{1} \otimes \mathbb{1}  + g_a \tau^a \otimes \tau^a + g_{4} \mathbb{1}\otimes \tau^3 \right)_{\alpha \beta, {\alpha '} {\beta '}},\label{ext:G_tensor}\\
	 \Lambda^{jl,km}  & = \left( \mathbb{1} \otimes \mathbb{1} - \sigma_z \otimes \sigma_z \right)^{jl,km}\label{ext:Lambda_tensor}
\end{align}
and the second, accounting for the terms proportional to $g_5$, is
\begin{align}
	\tilde{G}_{\alpha \beta, {\alpha '} {\beta '}} &= -g_{5}\left( \mathbb{1} \otimes \tau^3 \right)_{\alpha \beta, {\alpha '} {\beta '}},\label{ext:G_til_tensor}\\
	 \tilde{\Lambda}^{jl,km}  & = \left( \mathbb{1} \otimes \sigma_z - \sigma_z \otimes \mathbb{1} \right)^{jl,km}.\label{ext:Lambda_til_tensor}
\end{align}
Then, the system of equations describing the discontinuities of the wave-function $\phi(x_{1},x_{2})$ \eqref{DM:N_particle_state} extends the original system \eqref{DM:2p_syseq1}-\eqref{DM:2p_syseq4} as follows:
\begin{align} 
	i {\Delta^{(-)}}^{12}_{11} + 2 \tilde{\xi} (g_0 + g_3 + g_4) {\Delta^{(+)}}^{12}_{11} + 2 \tilde{\xi} (g_1 - g_2) {\Delta^{(+)}}^{12}_{22} &=0, \label{ext:syseq1}\\
	i {\Delta^{(-)}}^{12}_{12} + 2 \tilde{\xi} (g_0 - g_3-g_5) {\Delta^{(+)}}^{12}_{12} + 2 \tilde{\xi} (g_1 + g_2) {\Delta^{(+)}}^{12}_{21} &=0, \label{ext:syseq2} \\
	i {\Delta^{(-)}}^{12}_{21} + 2 \tilde{\xi} (g_0 - g_3 -g_5) {\Delta^{(+)}}^{12}_{21} + 2 \tilde{\xi} (g_1 + g_2) {\Delta^{(+)}}^{12}_{12} &=0,\label{ext:syseq3}\\
	i {\Delta^{(-)}}^{12}_{22} + 2 \tilde{\xi} (g_0 + g_3 - g_4) {\Delta^{(+)}}^{12}_{22} + 2 \tilde{\xi} (g_1 - g_2) {\Delta^{(+)}}^{12}_{11} &=0,\label{ext:syseq4}
\end{align}
in terms of the same shorthand combination of wave-functions \eqref{DM:2p_Delta}. Again, all the remaining combinations ${\Delta^{(\pm)}}^{ij}_{\alpha \beta}$ are continuous by virtue of \eqref{DM:2p_Delta_symmetry}.

To solve this system of equations, we first substitute the same wave function for the 2-particle sector obtained from the Bethe Ansatz for the original Belavin model \eqref{DM:2p_BA1} and \eqref{DM:2p_BA2}. The resulting system of equations for the $B$-amplitudes \eqref{DM:2p_B-amplitudes} reads:
\begin{align}
	&i \left(B_{0}^{12}+B_{0}^{21}\right)+\lambda_{0}\left(B_{0}^{12}-B_{0}^{21}\right)\coth(\theta)-\lambda_{5}
	\left(B_{3}^{12}+B_{3}^{21}\right) = 0,\label{ext:Bsystemeq3}\\
	&i\coth(\theta)\left(B_{1}^{21}-B_{1}^{12}\right) -\lambda_{1}\left(B_{1}^{21}+B_{1}^{12}\right) -\lambda_{4}\left(B_{2}^{21}+B_{2}^{12}\right) = 0,\label{ext:Bsystemeq1}\\
	&i\coth(\theta)\left(B_{2}^{21}- B_{2}^{12}\right) -\lambda_{2}\left(B_{2}^{21}+B_{2}^{12}\right) -\lambda_{4}\left(B_{1}^{21}+B_{1}^{12}\right) = 0,\label{ext:Bsystemeq2}\\
	&i\coth(\theta)\left(B_{3}^{21}-B_{3}^{12}\right) -\lambda_{3}\left(B_{3}^{21}+B_{3}^{12} \right) -\lambda_{5}\left(B_{0}^{12}-B_{0}^{21} \right)\coth(\theta) = 0.\label{ext:Bsystemeq4}
\end{align}
Here, we also use the same combinations of coupling constants introduced in \eqref{DM:2p_lambda1} - \eqref{DM:2p_lambda4} and similarly define $\lambda_{4}=2\tilde{\xi}g_{4}$ and $\lambda_{5}=2\tilde{\xi}g_{5}$. Thus, the $B$-amplitudes are related as:
\begin{align}\label{ext:2p_B_amp_general_relations}
	\begin{pmatrix}
		B_0^{12}\\
		B_1^{12}\\
		B_2^{12}\\
		B_3^{12}
	\end{pmatrix} = \begin{pmatrix}
		\alpha_1(\theta) & 0 & 0 & \delta_1(\theta)\\
		0 & \beta_1 \theta & \gamma_1(\theta) & 0 \\
		0 & \gamma_2(\theta) & \beta_2(\theta) & 0 \\
		\delta_2 (\theta) & 0 & 0 & \alpha_2(\theta) 
	\end{pmatrix} \begin{pmatrix}
		B_0^{21}\\
		B_1^{21}\\
		B_2^{21}\\
		B_3^{21}
	\end{pmatrix},
\end{align}
where the coefficients of the mixing matrix are given by:
\begin{align}
	\alpha_1(\theta) &= \frac{- i \lambda_3 + (1 + \lambda_0 \lambda_3 - \lambda_5^2) \coth \theta + i \lambda_0 \coth^2 \theta}{i \lambda_3 + (-1 + \lambda_0 \lambda_3 - \lambda_5^2) \coth \theta + i \lambda_0 \coth^2 \theta}, \label{ext:matrix_coefficients_a1}\\
	\alpha_2(\theta) &=   \frac{ -i \lambda_3 + (-1 - \lambda_0 \lambda_3 + \lambda_5^2) \coth \theta + i \lambda_0 \coth^2 \theta}{i \lambda_3 + (-1 + \lambda_0 \lambda_3 - \lambda_5^2) \coth \theta + i \lambda_0 \coth^2 \theta}, \label{ext:matrix_coefficients_a2} \\
	\beta_1(\theta) &= - \frac{\lambda_1 \lambda_2 - \lambda_4^2 + i (\lambda_1 - \lambda_2) \coth \theta + \coth^2 \theta}{\lambda_1 \lambda_2 - \lambda_4^2 + i (\lambda_1 + \lambda_2) \coth \theta - \coth^2 \theta}, \label{ext:matrix_coefficients_b1}\\
	\beta_2(\theta) &=  \frac{ - \lambda_1 \lambda_2 + \lambda_4^2 + i (\lambda_1 - \lambda_2) \coth \theta - \coth^2 \theta}{\lambda_1 \lambda_2 - \lambda_4^2 + i (\lambda_1 + \lambda_2) \coth \theta - \coth^2 \theta}, \label{ext:matrix_coefficients_b2}\\
	\gamma_1 (\theta) &= \gamma_2 (\theta) = \frac{2 i \lambda_4 \coth \theta}{- \lambda_1 \lambda_2 + \lambda_4^2 -i(\lambda_1 + \lambda_2) \coth \theta + \coth^2 \theta}, \label{ext:matrix_coefficients_c1_c2}\\
	\delta_1(\theta) &= - \delta_2(\theta) = - \frac{2 i \lambda_5 \coth \theta}{i \lambda_3 + (-1 + \lambda_0 \lambda_3 + \lambda_5^2) \coth \theta + i \lambda_0 \coth^2 \theta}. \label{ext:matrix_coefficients_d1_d2}
\end{align}
Note that, differently from the original Belavin model, in which the mixing matrix was diagonal (c.f. \eqref{DM:2p_B_amp_general_relations}), in the extende model, there is some mixing between different amplitudes due to the new interaction terms. 

We conclude this section by giving the explicit form of the Lagrangian \eqref{ext:Lagrangian_g4g5} written in terms of the  fields convenient for the purposes of considering the low-energy limit:
\begin{align}
	\chi^{\alpha}&:=-\frac{1}{2}\left(\psi^{\alpha}_{1}+i \psi^{\alpha}_{2}\right),\notag \\
	\zeta^{\alpha}&:=\frac{1}{2}\left(\psi^{\alpha}_{1}-i\psi^{\alpha}_{2}\right),\label{Bax:chi_fields}
\end{align}
The  Lagrangian \eqref{ext:Lagrangian_g4g5} in terms of these fields becomes:
\begin{align}
	\mathscr{L} =-&\frac{i}{2}\left( \chi^{\dagger \alpha} \partial_{0} \chi^{\alpha} + \zeta^{\dagger \alpha} \partial_{0} \zeta^{\alpha}\right) - \frac{1}{2} \left( \chi^{\dagger \alpha} \partial_{1} \zeta^{\alpha} - \zeta^{\dagger \alpha} \partial_{1} \chi^{\alpha} \right) \notag \\
	&+ \frac{m}{2}\left( \chi^{\dagger \alpha}\chi^{\alpha} - \zeta^{\dagger \alpha}\zeta^{\alpha}\right) - \mathscr{L}_{{int}},\label{ext:Lagrangian_density}
\end{align}
where the interaction term has the form:
\begin{align}\label{ext:explicit_Lagrangian_density}
	\mathscr{L}_{{int}} &= \left (\frac{g_{0}}{4}\right)
\left[
	\zeta^{\dagger 1} \zeta^{\dagger 2}\zeta^{1}\zeta^{2}
	+\chi^{\dagger 1}\chi^{\dagger 2}\zeta^{1}\zeta^{2}
	+2 \zeta^{\dagger 1} \chi^{\dagger 1}\zeta^{1}\chi^{1}
	+\zeta^{\dagger 1} \chi^{\dagger 2}\zeta^{1}\chi^{2}
	+\zeta^{\dagger 2} \chi^{\dagger 1}\zeta^{1}\chi^{2} \right.\notag\\
&\left. +2 \zeta^{\dagger 2} \chi^{\dagger 2}\zeta^{2}\chi^{2}
	+\zeta^{\dagger 1} \chi^{\dagger 2}\zeta^{2}\chi^{1}
	+\zeta^{\dagger 2} \chi^{\dagger 1}\zeta^{2}\chi^{1}
	+\zeta^{\dagger 1} \zeta^{\dagger 2}\chi^{1}\chi^{2}
	+\chi^{\dagger 1} \chi^{\dagger 2}\chi^{1}\chi^{2}
\right] \notag \displaybreak[3] \\
&+\left(\frac{g_{1}}{4}\right)
\left[
	-\zeta^{\dagger 1} \zeta^{\dagger 2}\zeta^{1}\zeta^{2}
	- \chi^{\dagger 1} \chi^{\dagger 2}\zeta^{1}\zeta^{2}
	+2 \zeta^{\dagger 2} \chi^{\dagger 2}\zeta^{1}\chi^{1}
	+\zeta^{\dagger 1} \chi^{\dagger 2}\zeta^{1}\chi^{2} 
	+\zeta^{\dagger 2} \chi^{\dagger 1}\zeta^{1}\chi^{2} \right.\notag \displaybreak[3] \\
&\left.	
	+\zeta^{\dagger 1} \chi^{\dagger 2}\zeta^{2}\chi^{1}
	+ \zeta^{\dagger 2} \chi^{\dagger 1}\zeta^{2}\chi^{1}
	+2 \zeta^{\dagger 1} \chi^{\dagger 1}\zeta^{2}\chi^{2}
	- \zeta^{\dagger 1} \zeta^{\dagger 2}\chi^{1}\chi^{2}
	- \chi^{\dagger 1} \chi^{\dagger 2}\chi^{1}\chi^{2}
\right] \notag \displaybreak[3] \\
&-\left(\frac{g_{2}}{4}\right)
\left[
	 \zeta^{\dagger 1} \zeta^{\dagger 2}\zeta^{1}\zeta^{2}
	+ \chi^{\dagger 1} \chi^{\dagger 2}\zeta^{1}\zeta^{2}
	+2 \zeta^{\dagger 2} \chi^{\dagger 2}\zeta^{1}\chi^{1}
	-\zeta^{\dagger 1} \chi^{\dagger 2}\zeta^{1}\chi^{2} 
	-\zeta^{\dagger 2} \chi^{\dagger 1}\zeta^{1}\chi^{2} \right.\notag \displaybreak[3] \\
&\left.	
	-\zeta^{\dagger 1} \chi^{\dagger 2}\zeta^{2}\chi^{1}
	-\zeta^{\dagger 2} \chi^{\dagger 1}\zeta^{2}\chi^{1}
	+2 \zeta^{\dagger 1} \chi^{\dagger 1}\zeta^{2}\chi^{2}
	+\zeta^{\dagger 1} \zeta^{\dagger 2}\chi^{1}\chi^{2}
	+\chi^{\dagger 1} \chi^{\dagger 2}\chi^{1}\chi^{2}
\right]. \notag \displaybreak[3] \\
&-\left(\frac{g_{3}}{4}\right)
\left[
	\zeta^{\dagger 1} \zeta^{\dagger 2}\zeta^{1}\zeta^{2}
	+\chi^{\dagger 1} \chi^{\dagger 2}\zeta^{1}\zeta^{2}
	-2\zeta^{\dagger 1} \chi^{\dagger 1}\zeta^{1}\chi^{1}
	+\zeta^{\dagger 1} \chi^{\dagger 2}\zeta^{1}\chi^{2} 
	+\zeta^{\dagger 2} \chi^{\dagger 1}\zeta^{1}\chi^{2} \right.\notag \displaybreak[3] \\
&\left.	
	+\zeta^{\dagger 1} \chi^{\dagger 2}\zeta^{2}\chi^{1}
	+\zeta^{\dagger 2} \chi^{\dagger 1}\zeta^{2}\chi^{1}
	-2\zeta^{\dagger 2} \chi^{\dagger 2}\zeta^{2}\chi^{2}
	+\zeta^{\dagger 1} \zeta^{\dagger 2}\chi^{1}\chi^{2}
	+\chi^{\dagger 1} \chi^{\dagger 2}\chi^{1}\chi^{2}
\right]. \notag \displaybreak[3] \\
&+\left(\frac{g_{4}}{2}\right)
	\left[
		\zeta^{\dagger 1} \chi^{\dagger 1}\zeta^{1}\chi^{1}
		-\zeta^{\dagger 2} \chi^{\dagger 2}\zeta^{2}\chi^{2}		
	\right]\notag \displaybreak[3] \\
&+\left(\frac{ig_{5}}{2}\right)
	\left[
		\zeta^{\dagger 1} \chi^{\dagger 2}\zeta^{1}\zeta^{2}
		+\zeta^{\dagger 2} \chi^{\dagger 1}\zeta^{1}\zeta^{2}
		-\zeta^{\dagger 1} \zeta^{\dagger 2}\zeta^{1}\chi^{2}
		-\chi^{\dagger 1} \chi^{\dagger 2}\zeta^{1}\chi^{2} 
		-\zeta^{\dagger 1} \zeta^{\dagger 2}\zeta^{2}\chi^{1}\right. \notag \displaybreak[3] \\
&\left. 
		-\chi^{\dagger 1} \chi^{\dagger 2}\zeta^{2}\chi^{1}
		+\zeta^{\dagger 1} \chi^{\dagger 2}\chi^{1}\chi^{2}
		+\zeta^{\dagger 2} \chi^{\dagger 1}\chi^{1}\chi^{2}
	\right].
\end{align}

\section{$S$-matrix factorization, Baxter's eight-vertex model and its exceptional solutiions}
\label{section:8vertex}
In this section, we analyze the problem of $S$-matrix factorization for $n \geq 3$, and its relation to the (generalized) eight-vertex model \cite{Baxter:1982zz} and in particular its exceptional solutions \cite{Khachatryan:2012wy}. We fix our notations following \cite{Belavin:1979pq,Dutyshev:1980vn,Samaj:2013yva}. Let $I=(1,2, \ldots, N)$ with $x_{1} < x_{2}< \ldots <x_{N}$ denote the fundamental ordering, and $P=(P_{1},P_{2}, \ldots,  P_{N})$ with  $x_{P_{1}}<x_{P_{2}}< \ldots,<x_{P_{N}}$ be an arbitrary ordering sector. Then the Bethe Ansatz for the wave-function in the $P$-sector, denoted by $\phi^{P}_{\alpha_{q_1}\cdots \alpha_{q_N}}(x)$, is of the form:
\begin{align}
	\phi^{P}_{\alpha_{1}\cdots \alpha_{N}}(x) = \sum\limits_{Q}\: \sign(Q)\, A^{PQ}_{\alpha_{P_{1}}, \ldots, \alpha_{P_{N}}}\prod\limits_{j=1}^{N} u_{j}\left(\theta_{Q_{j}}\right) \exp\left(i k_{Q_{j}}x_{j}\right),\label{8v:wave_function}
\end{align}
where $PQ$ stands for the product of permutations, $k_{i}=m\sinh(\theta_{i})$, and $u_{j}(\theta)=\frac{1}{\sqrt{2 \cosh(\theta)}}\begin{pmatrix}
           e^{-\nicefrac{\theta}{2}} \\
           e^{\nicefrac{\theta}{2}} \\
		   \end{pmatrix}$ is the spinor for the  $j{\textrm{\small th}}$ particle. This form of the wave-function satisfies the anti-symmetry condition \eqref{DM:antisymmetry_condition}. The quantum integrability of the model, expressed as the factorization of the $N$-particle  $S$-matrix in terms of two-particle $S$-matrices, follows from \eqref{8v:wave_function}, provided the Yang-Baxter equation is satisfied. This can be easily seen by recasting the above expression into the following equivalent form \cite{Samaj:2013yva}:
\begin{align}
	\phi^{P}(\sigma,x) = \sum\limits_{Q}\: \sign(P)\sign(Q)\, \mathcal{A}_{\sigma_{P_{1}}\sigma_{P_{2}} \cdots \sigma_{P_{N}}}(k_{Q_{1}},k_{Q_{2}},\ldots,k_{Q_{N}}) \exp\left(i \sum\limits_{j=1}^{N} k_{Q_{j}}x_{P_{j}}\right),\label{8v:wave_function_alternative_form}
\end{align}
where we have combined the spinor and isotropic indices into $\sigma_{i} = \{m_{i},\alpha_{i}\}$, and $(\sigma,x)$ stands for $(\sigma_{1},x_{1}; \ldots, \sigma_{N},x_{N})$.\footnote{The equivalence between \eqref{8v:wave_function} and \eqref{8v:wave_function_alternative_form} easily follows from identification:
\begin{align}\label{8v:identification}
	\mathcal{A}_{\sigma_{P_{1}}\sigma_{P_{2}} \cdots \sigma_{P_{N}}}(k_{Q_{1}},k_{Q_{2}},\ldots,k_{Q_{N}})=A^{Q}_{\alpha_{P_{1}},\ldots, \alpha_{P_{N}}}\prod\limits_{j=1}^{N} u^{m_{P_{j}}}\left(\theta_{Q_{j}}\right)
\end{align}
}
The $\mathcal{A}$-coefficients relate to the two-particle $S$-matrix via the exchange relation:
\begin{align}\label{8v:smatrix_def}
	\mathcal{A}_{\sigma_j \sigma_{i}}(k_{r},k_{s})=\sum\limits_{\sigma'_{i} \sigma'_{j}} S^{\sigma_{i} \sigma_{j}}_{\sigma'_{i} \sigma'_{j}}\mathcal{A}_{\sigma'_i \sigma'_{j}}(k_{s},k_{r}),
\end{align}
where $\{(i,j),\,(r,s)\} =\{(1,2),\,(2,1)\}$. The $S$-matrix should satisfy the standard normalization and unitarity conditions.

The factorization property of the $N$-particle scattering matrix readily follows, and the consistency for $N\ge 3$ requires the Yang-Baxter equation ($YBE$):
\begin{align}\label{8v:YBE}
	S_{12}(k_{1},k_{2})S_{13}(k_{1},k_{3})S_{23}(k_{2},k_{3})=S_{23}(k_{2},k_{3})S_{13}(k_{1},k_{3})S_{12}(k_{1},k_{2}),
\end{align}
to be satisfied. For all the cases considered in this paper the $S$-matrix elements depend only on the differences of rapidities: $\theta_{ij}:=\nicefrac{(\theta_{i}-\theta_{j})}{2}$. In this case the YBE equation has the form:
\begin{align}
\label{8v:YBE_rapidities}
	S_{12}(\theta_{12})S_{13}(\theta_{13})S_{23}(\theta_{23})=S_{23}(\theta_{23})S_{13}(\theta_{13})S_{12}(\theta_{12}).
\end{align}
For the original model the $S$-matrix has the form:
\begin{equation}\label{8v:Smatrix_original}
   S =  \begin{pmatrix}
       a(\theta) & 0 & 0 & d(\theta) \\
        0 & b(\theta) & c(\theta) & 0 \\
       0 & c(\theta) & b(\theta) & 0 \\
	   d(\theta) & 0 & 0 & a(\theta)
     \end{pmatrix}.
   \end{equation}
Thus, it has the form of the $R$-matrix for the eight-vertex model \cite{Baxter:1982zz}. The $YBE$ was investigated in \cite{Belavin:1979pq,Dutyshev:1980vn} for the original model and it was shown that it corresponds to Baxter's general solution in terms of the elliptic functions \cite{Baxter:1982zz} only in the massless case: $m=0$. In the next section, we analyze this point more carefully, and show that there exists an exceptional solution to $YBE$ for the massive case. Then, in the subsequent section we extend this result for the model including the interaction terms with $g_{4}$ and $g_{5}$ coupling constants, in which case the $S$-matrix is not of the type as in \eqref{8v:Smatrix_original}, but has a more general (inhomogeneous) form. The classification of $R$-matrices with inhomogeneous structure, together with the exceptional solutions to $YBE$, was considered in \cite{Khachatryan:2012wy, Hietarinta:1991ti}. We show that the integrability of the model with $g_{4}$ and $g_{5}$ coupling constants corresponds exactly to such exceptional solutions. 

\subsection{Baxter's general and exceptional solutions for homogeneous $S$-matrices}
\label{subsection:Bax}

It was shown by Baxter \cite{Baxter:1982zz} that the most general solution given in terms of elliptic functions for the $R$-matrix of the eight-vertex model, which has the same form as the $S$-matrix in \eqref{8v:Smatrix_original}, requires the following ratios to be constants:
\begin{align}
	\Gamma:= \frac{ab-cd}{ab+cd}, \quad \Delta:= \frac{a^2 +b^2 - c^2 -d^2}{2(ab+cd)}.\label{Bax:Gamma_Delta}
\end{align}
In this case, for the original model, the coefficients $(a,b,c,d)$ in \eqref{8v:Smatrix_original} are given by the following formulas:
\begin{align}
	a(\theta)&=\frac{\gamma_{1}(\theta)+\gamma_{2}(\theta)}{2},\label{Bax:a}\\ b(\theta)&=\frac{\gamma_{3}(\theta)-\gamma_{0}(\theta)}{2},\label{Bax:b} \\
	c(\theta)&=\frac{\gamma_{0}(\theta)+\gamma_{3}(\theta)}{2},\label{Bax:c}\\ d(\theta)&=\frac{\gamma_{1}(\theta)-\gamma_{2}(\theta)}{2},\label{Bax:d}
\end{align}
with  $\gamma_{\mu} (\theta)$, $0 \leq \mu \leq 3$ given by \eqref{DM:2p_gamma_coefs}.

The key point is that Baxter's condition \eqref{Bax:Gamma_Delta} and consequently the general solution in terms of elliptic functions are valid provided all six Yang-Baxter equations are independent \cite{Baxter:1982zz,Samaj:2013yva}. This indeed happens, as was shown in \cite{Belavin:1979pq,Dutyshev:1980vn}, only in the massless $(m=0)$ case. As a consequence, this requires taking the limit $\theta \rightarrow \infty$, in order to keep the momentum $p=m\sinh(\theta)$ finite. In such a limit, the formulas \eqref{Bax:a}-\eqref{Bax:d} simplify considerably together with the six independent Yang-Baxter equations, and the massless case can be reduced to Baxter's general solution for the eight-vertex model. 

We emphasize that, in general, verifying only the conditions \eqref{Bax:Gamma_Delta} is not enough. Namely, one should also verify that the six Yang-Baxter equations are independent. This is indeed the case for the coefficients $(a,b,c,d)$ in \eqref{Bax:a}-\eqref{Bax:d}. Even though, the parameters $\Gamma$ and $\Delta$ defined in \eqref{Bax:Gamma_Delta} are constants for some relations between $\lambda_{0},\ldots,\lambda_{3}$, they do not provide a solution for the Yang-Baxter equations, since the $YBE$ system fail to be independent. There are, however, some exceptional solutions to the $YBE$ which correspond to two possibilities: either some of the $(a,b,c,d)$ coefficients in \eqref{8v:Smatrix_original} coincide, or some of  the $(a,b,c,d)$ coefficients in \eqref{8v:Smatrix_original} are equal to zero. For a detailed analysis and a subsequent classification of such exceptional cases, we refer to the original paper \cite{Khachatryan:2012wy}. For the massive case $(m \neq 0)$, we find exactly such exceptional solutions, which we list below. First, we address the finiteness of the $S$-matrix and renormalize the coupling constants (see the comment after \eqref{DM:2p_lambda4}), and recall the relation between the original coupling constants $g_{0},\ldots,g_{3}$ and $\lambda_{\mu} = 2 \xi g_{\mu};\: \mu=0,\ldots,3$. In order to write the $S$-matrix in terms of finite quantities, one should start with the Lagrangian written in terms of the renormalized coupling constants:
\begin{equation}
	\tilde{g}_{\mu}:=\frac{g_{\mu}}{2 \xi}.\label{Bax:constant_renormalization}
\end{equation}
Below we list the (real) exceptional solutions  in terms of both $\lambda_{\mu}$ and $\tilde{g}_{\mu}$, and give, for each solution, the corresponding interaction Lagrangian, which can be readily found from the   explicit form of the general Lagrangian given in  \eqref{ext:explicit_Lagrangian_density}.

\paragraph{\bf Solution 1:} $\{\lambda_{1}=0, \,\lambda_{2}=0,\,\lambda_{3}=\lambda_{0}\} \Longleftrightarrow \{\tilde{g}_{1}=0,\,\tilde{g}_{2}=0,\,\tilde{g}_{3}=- \tilde{g}_{0}\}$.\\
In this case, the $S$-matrix has the form:
\begin{equation}\label{8v:Smatrix_solution1}
   S_{\textrm{(1)}} = \begin{pmatrix}
       1 & 0 & 0 & 0 \\
        0 & -\frac{\left(1+4\tilde{g}_{0}^{2} \right)\coth \theta}{\left(2\tilde{g}_{0} +i\coth \theta \right)\left(i +2\tilde{g}_{0}\coth \theta \right)} & \frac{2 \tilde{g}_{0}}{\left(-2i\tilde{g}_{0}\cosh \theta  + \sinh \theta \right) \left(i\cosh \theta +2\tilde{g}_{0}\sinh \theta  \right)} & 0 \\
       0 &\frac{2 \tilde{g}_{0}}{\left(-2i\tilde{g}_{0}\cosh \theta  + \sinh \theta \right) \left(i\cosh \theta +2\tilde{g}_{0}\sinh \theta  \right)} &  -\frac{\left(1+4\tilde{g}_{0}^{2} \right)\coth \theta}{\left(2\tilde{g}_{0} +i\coth \theta \right)\left(i +2\tilde{g}_{0}\coth \theta \right)} & 0 \\
	   0 & 0 & 0 & 1
     \end{pmatrix}.
   \end{equation}
\newline
The resulting interaction Lagrangian has the form:
\begin{align}\label{ext:explicit_Lagrangian_density_Sol1}
	\mathscr{L}_{{int}}^{\textrm{(1)}} &= \left (\frac{g_{0}}{2}\right)
\left[
	\zeta^{\dagger 1} \zeta^{\dagger 2}\zeta^{1}\zeta^{2}
	+\chi^{\dagger 1}\chi^{\dagger 2}\zeta^{1}\zeta^{2}
	+\zeta^{\dagger 1} \chi^{\dagger 2}\zeta^{1}\chi^{2}
	+\zeta^{\dagger 2} \chi^{\dagger 1}\zeta^{1}\chi^{2}
	+\zeta^{\dagger 1} \chi^{\dagger 2}\zeta^{2}\chi^{1} \right.\notag\\
&\left. 
	+ \zeta^{\dagger 2} \chi^{\dagger 1}\zeta^{2}\chi^{1}
	+\zeta^{\dagger 1} \zeta^{\dagger 2}\chi^{1}\chi^{2}
	+\chi^{\dagger 1} \chi^{\dagger 2}\chi^{1}\chi^{2}
\right].
\end{align}

\paragraph{\bf Solution 2:}\label{ext:Solution2} $\{\lambda_{1}=0, \,\lambda_{2}=\lambda_{0}, \,\lambda_{3}=0\} \Longleftrightarrow \{\tilde{g}_{1}=-\tilde{g}_0, \,\tilde{g}_{2}=0, \,\tilde{g}_{3}=0\}$.\\
In this case, the $S$-matrix has the form:
\begin{equation}\label{8v:Smatrix_solution2}
   S_{\textrm{(2)}} = \begin{pmatrix}
      \frac{i\coth \theta }{2\tilde{g}_{0} +i \coth \theta } & 0 & 0 & \frac{2\tilde{g}_{0}}{2\tilde{g}_{0} +i \coth \theta} \\
        0 & \frac{i}{i + 2\tilde{g}_{0}\coth \theta } & \frac{2\tilde{g}_{0}\coth \theta }{i + 2\tilde{g}_{0}\coth \theta } & 0 \\
        0 & \frac{2\tilde{g}_{0}\coth \theta }{i + 2\tilde{g}_{0}\coth \theta } & \frac{i}{i + 2\tilde{g}_{0}\coth \theta } & 0 \\
	   \frac{2\tilde{g}_{0}}{2\tilde{g}_{0} +i \coth \theta } & 0 & 0 &  \frac{i\coth \theta }{2\tilde{g}_{0} +i \coth \theta }
     \end{pmatrix}.
   \end{equation}
The resulting interaction Lagrangian has the form:
\begin{align}\label{ext:explicit_Lagrangian_density_Sol2}
	\mathscr{L}_{{int}}^{\textrm{(2)}} &= \left (\frac{g_{0}}{4}\right)
\left[
	\zeta^{\dagger 1} \zeta^{\dagger 2}\zeta^{1}\zeta^{2}
	+\chi^{\dagger 1}\chi^{\dagger 2}\zeta^{1}\zeta^{2}
	+4\zeta^{\dagger 1} \chi^{\dagger 1}\zeta^{1}\chi^{2}
	-2\zeta^{\dagger 2} \chi^{\dagger 2}\zeta^{1}\chi^{1}
	-\zeta^{\dagger 1} \chi^{\dagger 2}\zeta^{1}\chi^{2}
	- \zeta^{\dagger 2} \chi^{\dagger 1}\zeta^{1}\chi^{2} \right.\notag\\
&\left. 
	-\zeta^{\dagger 1} \chi^{\dagger 2}\zeta^{2}\chi^{1}
	-\zeta^{\dagger 2} \chi^{\dagger 1}\zeta^{2}\chi^{1}
	-2\zeta^{\dagger 1} \chi^{\dagger 1}\zeta^{2}\chi^{2}
	+4\zeta^{\dagger 2} \chi^{\dagger 2}\zeta^{2}\chi^{2}
	+\zeta^{\dagger 1} \zeta^{\dagger 2}\chi^{1}\chi^{2}
	+\chi^{\dagger 1} \chi^{\dagger 2}\chi^{1}\chi^{2}
\right].
\end{align}

\paragraph{\bf Solution 3:} $\{\lambda_{1}=\lambda_{0}, \,\lambda_{2}=0, \,\lambda_{3}=0\} \Longleftrightarrow \{\tilde{g}_{1}=0, \,\tilde{g}_{2}=-\tilde{g}_{0}, \,\tilde{g}_{3}=0\}$.\\
In this case, the $S$-matrix has a same form similar to that of the Solution 2:
\begin{equation}\label{8v:Smatrix_solution3}
   S_{\textrm{(3)}} = \begin{pmatrix}
      \frac{i\coth \theta }{2\tilde{g}_{0} +i \coth \theta } & 0 & 0 & -\frac{2\tilde{g}_{0}}{2\tilde{g}_{0} +i \coth \theta } \\
        0 & \frac{i}{i + 2\tilde{g}_{0}\coth \theta } & \frac{2\tilde{g}_{0}\coth \theta }{i + 2\tilde{g}_{0}\coth \theta } & 0 \\
       0 & \frac{2\tilde{g}_{0}\coth \theta }{i + 2\tilde{g}_{0}\coth \theta } & \frac{i}{i + 2\tilde{g}_{0}\coth \theta } & 0 \\
	   - \frac{2\tilde{g}_{0}}{2\tilde{g}_{0} +i \coth \theta } & 0 & 0 &  \frac{i\coth \theta }{2\tilde{g}_{0} +i \coth \theta}
     \end{pmatrix}.
   \end{equation}
The resulting interaction Lagrangian has the form:
\begin{align}\label{ext:explicit_Lagrangian_density_Sol3}
	\mathscr{L}_{{int}}^{\textrm{(3)}} &= \left (\frac{g_{0}}{2}\right)
\left[
	\zeta^{\dagger 1} \zeta^{\dagger 2}\zeta^{1}\zeta^{2}
	+\chi^{\dagger 1}\chi^{\dagger 2}\zeta^{1}\zeta^{2}
	+\zeta^{\dagger 1} \chi^{\dagger 1}\zeta^{1}\chi^{1}
	+\zeta^{\dagger 2} \chi^{\dagger 2}\zeta^{1}\chi^{2}
	+\zeta^{\dagger 1} \chi^{\dagger 1}\zeta^{2}\chi^{2} \right.\notag\\
&\left. 
	+\zeta^{\dagger 2} \chi^{\dagger 2}\zeta^{2}\chi^{2}
	+\zeta^{\dagger 1} \zeta^{\dagger 2}\chi^{1}\chi^{2}
	+\chi^{\dagger 1} \chi^{\dagger 2}\chi^{1}\chi^{2}
\right].
\end{align}

\paragraph{\bf Solution 4:} $\{\lambda_{1}=\lambda_{2}=\lambda_{3}=-\nicefrac{1}{\lambda_{0}}\} \Longleftrightarrow \{\tilde{g}_{1,2,3}=\frac{1}{3}\left(-g_{0} \pm \sqrt{3+4\tilde{g}_{0}^{2}}\right)\}$.\\
In this case, the $S$-matrix can be written in the form:
\begin{equation}\label{8v:Smatrix_solution4}
   S_{\textrm{(4)}} = -\frac{-2\tilde{g}_{0}+\sqrt{3+4\tilde{g}_{0}^{2}}\pm 3i\coth \theta}{-2\tilde{g}_{0}+\sqrt{3+4\tilde{g}_{0}^{2}}\mp 3i\coth \theta }\mathscr{P},
   \end{equation}
where $\mathscr{P}$ is the permutation matrix:
\begin{align}\label{8v:permutation_matrix}
	\mathscr{P}=\begin{pmatrix}
	       1 & 0 & 0 & 0 \\
	        0 & 0 & 1 & 0 \\
	       0 & 1 & 0 & 0 \\
		   0 & 0 & 0 & 1
	     \end{pmatrix}.
\end{align}
It is interesting to note that this solution imposes the following  bound on the coupling constants:
\begin{equation}
	\tilde{g}_{1,2,3} \geq \nicefrac{1}{2}.\label{8v:bound}
\end{equation}
There are two resulting interaction Lagrangians corresponding to $(\pm)$ solutions above:
\begin{align}\label{ext:explicit_Lagrangian_density_Sol4a}
	\mathscr{L}_{{int}}^{\textrm{(4\,a)}} &= \left (\frac{g_{0}}{12}\right)
\left[
	6\zeta^{\dagger 1} \zeta^{\dagger 2}\zeta^{1}\zeta^{2}
	+6\chi^{\dagger 1}\chi^{\dagger 2}\zeta^{1}\zeta^{2}
	+4\zeta^{\dagger 1}\chi^{\dagger 1}\zeta^{1}\chi^{1}
	+2\zeta^{\dagger 1} \chi^{\dagger 2}\zeta^{1}\chi^{2}
	+2\zeta^{\dagger 2} \chi^{\dagger 1}\zeta^{1}\chi^{2}\right.\notag\\
&\left.
	+2\zeta^{\dagger 1} \chi^{\dagger 2}\zeta^{2}\chi^{1}
	+2\zeta^{\dagger 2} \chi^{\dagger 1}\zeta^{2}\chi^{1}
	+4\zeta^{\dagger 2} \chi^{\dagger 2}\zeta^{2}\chi^{2}
	+6\zeta^{\dagger 1} \zeta^{\dagger 2}\chi^{1}\chi^{2}
	+6\chi^{\dagger 1} \chi^{\dagger 2}\chi^{1}\chi^{2}
\right]\notag\\
&+\left (\frac{ \sqrt{3+4\tilde{g}_{0}^{2}}}{12}\right)
\left[
	-3\zeta^{\dagger 1} \zeta^{\dagger 2}\zeta^{1}\zeta^{2}
	-3\chi^{\dagger 1} \chi^{\dagger 2}\zeta^{1}\zeta^{2}
	+2\zeta^{\dagger 1} \chi^{\dagger 1}\zeta^{1}\chi^{1}
	+\zeta^{\dagger 1} \chi^{\dagger 2}\zeta^{1}\chi^{2}
	 \right.\notag\\
&\left.
	+\zeta^{\dagger 2} \chi^{\dagger 1}\zeta^{1}\chi^{2}
	+\zeta^{\dagger 1} \chi^{\dagger 2}\zeta^{2}\chi^{1}
	+\zeta^{\dagger 2} \chi^{\dagger 1}\zeta^{2}\chi^{1}
	+2\zeta^{\dagger 2} \chi^{\dagger 2}\zeta^{2}\chi^{2}
	-3\zeta^{\dagger 1} \zeta^{\dagger 2}\chi^{1}\chi^{2}
	-3\chi^{\dagger 1} \chi^{\dagger 2}\chi^{1}\chi^{2}
\right].
\end{align}   
and
\begin{align}\label{ext:explicit_Lagrangian_density_Sol4b}
	\mathscr{L}_{{int}}^{\textrm{(4\,b)}} &= \left (\frac{g_{0}}{12}\right)
\left[
	6\zeta^{\dagger 1} \zeta^{\dagger 2}\zeta^{1}\zeta^{2}
	+6\chi^{\dagger 1}\chi^{\dagger 2}\zeta^{1}\zeta^{2}
	+4\zeta^{\dagger 1}\chi^{\dagger 1}\zeta^{1}\chi^{1}
	+2\zeta^{\dagger 1} \chi^{\dagger 2}\zeta^{1}\chi^{2}
	+2\zeta^{\dagger 2} \chi^{\dagger 1}\zeta^{1}\chi^{2}\right.\notag\\
&\left.
	+2\zeta^{\dagger 1} \chi^{\dagger 2}\zeta^{2}\chi^{1}
	+2\zeta^{\dagger 2} \chi^{\dagger 1}\zeta^{2}\chi^{1}
	+4\zeta^{\dagger 2} \chi^{\dagger 2}\zeta^{2}\chi^{2}
	+6\zeta^{\dagger 1} \zeta^{\dagger 2}\chi^{1}\chi^{2}
	+6\chi^{\dagger 1} \chi^{\dagger 2}\chi^{1}\chi^{2}
\right]\notag\\
&+\left (\frac{ \sqrt{3+4\tilde{g}_{0}^{2}}}{12}\right)
\left[
	3\zeta^{\dagger 1} \zeta^{\dagger 2}\zeta^{1}\zeta^{2}
	+3\chi^{\dagger 1} \chi^{\dagger 2}\zeta^{1}\zeta^{2}
	-2\zeta^{\dagger 1} \chi^{\dagger 1}\zeta^{1}\chi^{1}
	-\zeta^{\dagger 1} \chi^{\dagger 2}\zeta^{1}\chi^{2}
	 \right.\notag\\
&\left.
	-\zeta^{\dagger 2} \chi^{\dagger 1}\zeta^{1}\chi^{2}
	-\zeta^{\dagger 1} \chi^{\dagger 2}\zeta^{2}\chi^{1}
	-\zeta^{\dagger 2} \chi^{\dagger 1}\zeta^{2}\chi^{1}
	-2\zeta^{\dagger 2} \chi^{\dagger 2}\zeta^{2}\chi^{2}
	+3\zeta^{\dagger 1} \zeta^{\dagger 2}\chi^{1}\chi^{2}
	+3\chi^{\dagger 1} \chi^{\dagger 2}\chi^{1}\chi^{2}
\right].
\end{align}
Note that the square root structure is somewhat similar to that of the $AAF$ model \eqref{ext:aaf_lagrangian}, for which, as we commented in section \ref{section:Extensions}, $S$-matrix factorization requires the condition $(g_{2})^{2}=g_{3}$ for the coupling constants in \eqref{ext:aaf_lagrangian}.

\paragraph{\bf Solution 5:} $\{\lambda_{0}=0, \,\lambda_{1}=\lambda_{2}, \,\lambda_{3}=0\} \Longleftrightarrow \{\tilde{g}_{1}=0, \,\tilde{g}_{2}=0, \,\tilde{g}_{3}=\tilde{g}_{0}\}$.\\
In this case, the $S$-matrix has the form:
\begin{equation}\label{8v:Smatrix_solution5}
   S_{\textrm{(5)}} = \begin{pmatrix}
	       -\frac{2\tilde{g}_{0}-i\coth \theta }{2\tilde{g}_{0}+i\coth \theta } & 0 & 0 & 0 \\
	        0 & 1 & 0 & 0 \\
	       0 & 0 & 1 & 0 \\
		   0 & 0 & 0 & -\frac{2\tilde{g}_{0}-i\coth \theta }{2\tilde{g}_{0}+i\coth \theta }
	     \end{pmatrix}.
   \end{equation}
The resulting interaction Lagrangian has the form:
\begin{align}\label{ext:explicit_Lagrangian_density_Sol5}
	\mathscr{L}_{{int}}^{\textrm{(5)}} &= \left (\frac{g_{0}}{2}\right)
\left[
	\zeta^{\dagger 1} \chi^{\dagger 1}\zeta^{1}\chi^{1}
	+\zeta^{\dagger 2}\chi^{\dagger 2}\zeta^{2}\chi^{2}
\right].
\end{align} 

In the next section we extend these considerations for the interaction terms with $g_{4}$ and $g_{5}$ coupling constants.

\subsection{Exceptional $YBE$ solutions for inhomogeneous $S$-matrices}
\label{subsection:inho}

The inclusion of the interaction terms proportional to the coupling contants $g_{4}$ and $g_{5}$ results in the following (inhomogeneous) $S$-matrix:
\begin{equation}\label{inho:Smatrix_g4_g5}
   S =  \begin{pmatrix}
       a_{1}(\theta) & 0 & 0 & d_{1}(\theta) \\
        0 & b_{1}(\theta) & c_{1}(\theta) & 0 \\
       0 & c_{2}(\theta) & b_{2}(\theta) & 0 \\
	   d_{2}(\theta) & 0 & 0 & a_{2}(\theta)
     \end{pmatrix}, 
   \end{equation}
where 
\begin{align}
	a_{1}(\theta)&= \frac{- \lambda_{1} \lambda_{2} + \lambda_{4}^{2} - 2 i \lambda_{4}\coth \theta - \coth^{2} \theta }{\lambda_{1} \lambda_{2} - \lambda_{4}^{2} +  i (\lambda_{1} +\lambda_{2})\coth \theta  - \coth^{2} \theta},\label{inho:a1}\\
	\notag\\
	a_{2}(\theta)&= \frac{-\lambda_{1} \lambda_{2} + \lambda_{4}^{2} +2 i \lambda_{4}\coth \theta  - \coth^{2} \theta }{\lambda_{1} \lambda_{2} - \lambda_{4}^{2} +  i (\lambda_{1} +\lambda_{2})\coth \theta  - \coth^{2} \theta},\label{inho:a2}\\
	\notag\\
	b_{1}(\theta)&= - \frac{\left[\lambda_{0} \lambda_{3} - (i -\lambda_{5})^{2} \right]\coth^{2} \theta }{i\lambda_{3} + \left[ -1 + \lambda_{0} \lambda_{3} - \lambda_{5}^{2} \right]\coth \theta  +i\lambda_{0}\coth^{2} \theta },\label{inho:b1}\\
	\notag\\
	b_{2}(\theta)&=-\frac{\left[\lambda_{0} \lambda_{3} - (i +\lambda_{5})^{2} \right]\coth^{2} \theta}{i\lambda_{3} + \left[ -1 + \lambda_{0} \lambda_{3} - \lambda_{5}^{2} \right]\coth \theta  +i\lambda_{0}\coth^{2} \theta},\label{inho:b2}\\
	\notag\\
	c_{1}(\theta)&=c_{2}(\theta) = - \frac{i\left(\lambda_{3} -\lambda_{0}\coth^{2} \theta \right)}{i\lambda_{3} + \left[-1 +\lambda_{0} \lambda_{3} -\lambda_{5}^{2}\right] \coth \theta  +i \lambda_{0}\coth^{2} \theta}, \label{inho:c1_c2}\\
	\notag\\
	d_{1}(\theta)&=d_{2}(\theta)= \frac{\left(\lambda_{2} - \lambda_{1}\right)\coth \theta }{-i\left[\lambda_{1}\lambda_{2} -\lambda_{4}^{2}\right] + \left(\lambda_{1} +\lambda_{2}\right)\coth \theta  +i  \coth^{2}\theta }.\label{inho:d1_d2}
\end{align}
The coefficients $(a_{i},b_{i},c_{i},d_{i})$, $i=1,2$ for the $S$-matrix can be trivially obtained from the coefficients of the mixing matrix $(\alpha_{i},\beta_{i},\gamma_{i},\delta_{i})$, $i=1,2$ \eqref{ext:matrix_coefficients_a1} - \eqref{ext:matrix_coefficients_d1_d2} by simply changing from the $B$-amplitudes back to the original $A$-amplitudes using \eqref{DM:2p_B-amplitudes}. Thus, we have $a_{1} \neq a_{2}$ and $b_{1} \neq b_{2}$. In such a case, there are in general twelve independent Yang-Baxter equations following from \eqref{8v:YBE_rapidities} (for a detailed exposition see \cite{Khachatryan:2012wy}):
\begin{align}\label{8v:YBB}
	&a_{1}(\theta)b_{1}(\theta+\theta')c_{1}(\theta')-b_{1}(\theta')c_{1}(\theta)c_{1}(\theta+\theta')-a_{1}(\theta+\theta')b_{1}(\theta)c_{1}(\theta')+
		\eta b_{2}(\theta')d_{1}(\theta)d_{1}(\theta+\theta')=0, \notag \displaybreak[3] \\
	&a_{1}(\theta)a_{1}(\theta')c_{1}(\theta+\theta')-b_{1}(\theta')b_{2}(\theta)c_{1}(\theta+\theta')-
		a_{1}(\theta+\theta')c_{1}(\theta)c_{1}(\theta')+ \eta a_{2}(\theta+\theta') d_{1}(\theta)d_{1}(\theta')=0,\notag \\
	&a_{1}(\theta')b_{2}(\theta+\theta')c_{1}(\theta)-a_{1}(\theta+\theta')b_{2}(\theta')c_{1}(\theta)-b_{2}(\theta)c_{1}(\theta+\theta')c_{1}(\theta')+
		\eta b_{1}(\theta)d_{1}(\theta+\theta')d_{1}(\theta')=0, \notag \displaybreak[3] \\
	&a_{2}(\theta+\theta')b_{1}(\theta')c_{1}(\theta)+b_{1}(\theta)c_{1}(\theta+\theta')c_{1}(\theta')-\eta b_{2}(\theta)d_{1}(\theta+\theta')d_{1}(\theta')-
		a_{2}(\theta')b_{1}(\theta+\theta')c_{1}(\theta)=0, \notag \displaybreak[3] \\
	&b_{1}(\theta)b_{2}(\theta')c_{1}(\theta+\theta')+a_{2}(\theta+\theta')c_{1}(\theta)c_{1}(\theta')-a_{2}(\theta)a_{2}(\theta')c_{1}(\theta+\theta')-
		\eta a_{1}(\theta+\theta')d_{1}(\theta)d_{1}(\theta')=0, \notag \displaybreak[3] \\
	&a_{2}(\theta)b_{2}(\theta+\theta')c_{1}(\theta')+\eta b_{1}(\theta')d_{1}(\theta)d_{1}(\theta+\theta')-b_{2}(\theta')c_{1}(\theta)c_{1}(\theta+\theta')-
		a_{2}(\theta+\theta')b_{2}(\theta)c_{1}(\theta')=0, \notag \displaybreak[3] \\
	&a_{2}(\theta')c_{1}(\theta+\theta')d_{1}(\theta)-a_{1}(\theta')c_{1}(\theta)d_{1}(\theta+\theta')+a_{1}(\theta)a_{1}(\theta+\theta')d_{1}(\theta')-
		b_{1}(\theta)b_{1}(\theta+\theta')d_{1}(\theta')=0, \notag  \displaybreak[3] \\
	&b_{2}(\theta+\theta')c_{1}(\theta')d_{1}(\theta)+a_{1}(\theta)b_{1}(\theta')d_{1}(\theta+\theta')-a_{1}(\theta')b_{2}(\theta)d_{1}(\theta+\theta')-
		b_{1}(\theta+\theta')c_{1}(\theta)d_{1}(\theta')=0, \notag  \displaybreak[3] \\
	&b_{2}(\theta+\theta')b_{2}(\theta')d_{1}(\theta)+a_{1}(\theta)c_{1}(\theta')d_{1}(\theta+\theta')-a_{2}(\theta)c_{1}(\theta+\theta')d_{1}(\theta')-
		a_{1}(\theta+\theta')a_{1}(\theta')d_{1}(\theta)=0, \notag \displaybreak[3] \\
	&a_{2}(\theta+\theta')a_{2}(\theta')d_{1}(\theta)-b_{1}(\theta+\theta')b_{1}(\theta')d_{1}(\theta)-a_{2}(\theta)c_{1}(\theta')d_{1}(\theta+\theta')+
		a_{1}(\theta)c_{1}(\theta+\theta')d_{1}(\theta')=0, \notag \displaybreak[3] \\
	&a_{2}(\theta')b_{1}(\theta)d_{1}(\theta+\theta')-a_{2}(\theta)b_{2}(\theta')d_{1}(\theta+\theta')+b_{2}(\theta+\theta')c_{1}(\theta)d_{1}(\theta')-
		b_{1}(\theta+\theta')c_{1}(\theta')d_{1}(\theta)=0, \notag \displaybreak[3] \\
	&a_{2}(\theta')c_{1}(\theta)d_{1}(\theta+\theta')-a_{2}(\theta)a_{2}(\theta+\theta')d_{1}(\theta')+b_{2}(\theta)b_{2}(\theta+\theta')d_{1}(\theta')-
		a_{1}(\theta')c_{1}(\theta+\theta')d_{1}(\theta)=0.
\end{align}
Here $\theta$ stands for $\theta_{12}$ and $\theta'$ stands for $\theta_{23}$ (c.f. \eqref{8v:YBE_rapidities}). The parameter $\eta$  is defined via the relation $d_{2}=\eta d_{1}$, and is required to be a constant for consistency of the Yang-Baxter equations.

A necessary condition for a general solution of the inhomogeneous $YBE$ to exist is  \cite{Khachatryan:2012wy}:
\begin{align}\label{inho:necessary_condition}
	\frac{b_1}{b_2} = \pm 1.
\end{align}
However, for the extended Belavin model, the $b$-coefficients satisfy:
\begin{align}
\frac{b_{1}}{b_{2}} = \frac{\lambda_{0} \lambda_{3} - (i -\lambda_{5})^{2}}{\lambda_{0} \lambda_{3} - (i +\lambda_{5})^{2}}.
\end{align} 
Hence, only for $\lambda_{5} = 0$ the condition \eqref{inho:necessary_condition} holds. Fixing $g_5 = 0$, there exists only one general real solution with non-zero $g_4$, corresponding to $\tilde{g}_{0}=\tilde{g}_{1}=\tilde{g}_{2}=\tilde{g}_{3}=0,\,\tilde{g}_{4} \neq 0$. Nonetheless, this solution is rather trivial, since it corresponds to a pair of decoupled Thirring models \eqref{ext:explicit_Lagrangian_density}. Besides this real solution, we note, for completeness sake, that it is possible to find some quite non-trivial solutions involving complex coupling constants, the physical meaning of which is currently not clear to us.

Nevertheless, since the general analysis conducted in \cite{Khachatryan:2012wy} relied on non-vanishing and non-coinciding $S$-matrix elements, it is still possible to find some exceptional solutions to the $YBE$ with $g_4\neq 0$ and $g_5 \neq 0$ by violating any of these hypotheses. One such solution corresponds to:
\begin{align}
	\lambda_{0}=\lambda_{3}=0;\:\:\lambda_{1}=\lambda_{2},\label{inho:l4_l5_general_sol}
\end{align}
or, in terms of the renormalized coupling constants:
\begin{align}
	\tilde{g}_{1}=\tilde{g}_{2}=0;\:\: \tilde{g}_{0}=\tilde{g}_{3},\label{inho:g4_g5_renorm_l5}
\end{align}
so that the coupling constants $\tilde{g}_{4}$ and $\tilde{g}_{5}$ are left unconstrained. The resulting $S$-matrix is:
\begin{equation}\label{inho:Smatrix_g4_g5_sol2}
   S_{\textrm{(6)}} =  \begin{pmatrix}
       -\frac{2\tilde{g}_{0}+\tilde{g}_{4} -i\coth \theta }{2\tilde{g}_{0}+\tilde{g}_{4} +i\coth \theta } & 0 & 0 & 0 \\
        0 & \frac{i -\tilde{g}_{5}}{i+\tilde{g}_{5}} & 0 & 0 \\
       0 & 0 & \frac{i +\tilde{g}_{5}}{i-\tilde{g}_{5}} & 0 \\
	   0 & 0 & 0 & -\frac{2\tilde{g}_{0}-\tilde{g}_{4} -i\coth \theta }{2\tilde{g}_{0}-\tilde{g}_{4} +i\coth \theta }
     \end{pmatrix}.
   \end{equation}
Note that in the limit $\tilde{g}_{4} \to 0$, the above $S$-matrix reduces to that of solution 5 \eqref{8v:Smatrix_solution5}.
The interaction Lagrangian corresponding to this case has the following form:
\begin{align}\label{ext:explicit_Lagrangian_density_Sol6}
	\mathscr{L}_{{int}}^{\textrm{(6)}} &= \left( g_{0} \right)
\left[
	\zeta^{\dagger 1} \chi^{\dagger 1}\zeta^{1}\chi^{1}
	+\zeta^{\dagger 2}\chi^{\dagger 2}\zeta^{2}\chi^{2}\right]
+\left(\frac{g_{4}}{2}\right)
\left[
	\zeta^{\dagger 1} \chi^{\dagger 1}\zeta^{1}\chi^{1}
	-\zeta^{\dagger 2}\chi^{\dagger 2}\zeta^{2}\chi^{2}
\right]\notag\\
&\left(\frac{i g_{5}}{2} \right)
\left[
	\zeta^{\dagger 1} \chi^{\dagger 2}\zeta^{1}\zeta^{2}
	+\zeta^{\dagger 2} \chi^{\dagger 1}\zeta^{1}\zeta^{2}
	-\zeta^{\dagger 1} \zeta^{\dagger 2}\zeta^{1}\chi^{2}
	-\chi^{\dagger 1} \chi^{\dagger 2}\zeta^{1}\zeta^{2}
	-\zeta^{\dagger 1} \zeta^{\dagger 2}\zeta^{2}\chi^{1} \right.\notag\\
&\left. 
	+\zeta^{\dagger 1} \chi^{\dagger 2}\chi^{1}\chi^{2}
	+\zeta^{\dagger 2} \chi^{\dagger 1}\chi^{1}\chi^{2}
 \right].
\end{align}

\section{Relating $XXZ$ and six-vertex models to exceptional solutions}
\label{section:xxz}

In this section, we consider in details some of the solutions previously derived in section \ref{subsection:Bax} and match their $S$-matrices to known models. To do so, we will first factorize each $S$-matrix as
\begin{align}\label{xxz:factorization}
	S_{(i)} = h_{(i)}{(\theta, \tilde{g}_0)} \tilde{S}_{(i)}, \quad i=1,2,3.
\end{align}
so that the $c$-coefficients of $\tilde{S}_{(i)}$ are normalized to one. Thereby, we will be able to identify the transformation of variables which relate the models described by the $S$-matrices of section \ref{subsection:Bax} to the $XXZ$ and six-vertex models. For the sake of completeness, before proceeding with this analysis, we recall the form of the $XYZ$ Hamiltonian on a closed chain of $N$ sites
\begin{align}\label{xyz:heisenberg_hamiltonian}
	\mathscr{H} = - \frac{1}{2} \sum_{n=1}^N \left( J_x \: \bm{\sigma}_x^{n} \bm{\sigma}_x^{n+1} + J_y \: \bm{\sigma}_y^{n} \bm{\sigma}_y^{n+1} + J_z \: \bm{\sigma}_z^{n} \bm{\sigma}_z^{n+1} \right),
\end{align}
where the spin-$\nicefrac{1}{2}$ operators are represented by Pauli spin operators.

\subsection{Solution 1}
The $S$-matrix for the six-vertex model on a square lattice can be parametrized in terms of hyperbolic functions as \cite{Baxter:1982zz}:
\begin{align}\label{xxz:baxter6}
	S_{(6V)} =  \begin{pmatrix}
		\rho \frac{\sinh(\lambda - u)}{\sinh(\lambda)} & 0 & 0 & 0\\
		0 & \rho \frac{\sinh(u)}{\sinh{\lambda}} & \rho & 0\\
		0 & \rho & \rho \frac{\sinh(u)}{\sinh{\lambda}} & 0\\
		0 & 0 & 0 & \rho \frac{\sinh(\lambda - u)}{\sinh(\lambda)} 
	\end{pmatrix},
\end{align}
with respect to the complex parameters $u, \lambda, \rho$. It can then be considered an entire function of $u$, while the remaining variables $\lambda$ and  $\rho$ are treated as constants, the latter being just some normalization factor usually taken as unit. The coupling constants $J_{x},J_{y},J_{z}$ of the Hamiltonian \eqref{xyz:heisenberg_hamiltonian} describing the corresponding $XXZ$ model can be obtained from the following relations (see, for example, \cite{Samaj:2013yva}):
\begin{align} \label{xxz:6J}
	J_{x} &= 1+ \tanh^2 \lambda \quad \text{and} \quad J_{y}= J_{z} = \sech^2 \lambda. 
\end{align}

To match the $S$-matrix \eqref{8v:Smatrix_solution1} describing solution 1: $\left(\tilde{g}_{1}=0, \tilde{g}_{2}=0, \tilde{g}_{3}=- \tilde{g}_{0}\right)$, we introduce, according to \eqref{xxz:factorization},
\begin{align}\label{xxz:h1}
	h_{(1)} (\theta, \tilde{g}_0) = \frac{4 \tilde{g}_0}{4 \tilde{g}_0 \cosh 2 \theta + i \left( 1 - 4 \tilde{g}_0^2 \right) \sinh 2 \theta}.
\end{align}
In this case, the resulting $S$-matrix 
\begin{align}
	\tilde{S}_{(1)} = \begin{pmatrix}
		\cosh 2 \theta + \frac{i\left( 1 - 4 \tilde{g}_0^2 \right) \sinh 2 \theta}{4 \tilde{g}_0}  & 0 & 0 & 0\\
		0 & \frac{i \left( 1 + 4 \tilde{g}_0^2 \right) \sinh 2 \theta}{4 \tilde{g_0}}  & 1 & 0\\
		0 & 1 & \frac{i \left( 1 + 4 \tilde{g}_0^2\right)\sinh 2 \theta}{4 \tilde{g_0}} & 0 \\
		0 & 0 & 0 & \cosh 2 \theta + \frac{i\left( 1 - 4 \tilde{g}_0^2 \right)\sinh 2 \theta}{4 \tilde{g}_0} 
	\end{pmatrix}
\end{align}
can be reduced to Baxter's six-vertex $S$ matrix \eqref{xxz:baxter6} provided we take the rapidity $\theta = \nicefrac{u}{2}$, fix $\rho=1$ and identify the parameter $\lambda$ as:
\begin{align}
	\lambda =  i \cot^{-1} \left[ \frac{1}{4\tilde{g}_0} - \tilde{g}_0 \right]
\end{align}
with $| \tilde{g}_0 | \geq \nicefrac{1}{2}$. Therefore, according to \eqref{xxz:6J}, solution 1 corresponds to a $XXZ$ model described by
\begin{align}\label{xxz:JS1}
	J_x = 1 - \frac{16 \tilde{g}_0^2}{\left( 1 - 4 \tilde{g}_0^2\right)^2} \quad  \text{and} \quad J_y = J_z = \left( \frac{1 + 4 \tilde{g}_0^2}{1 - 4 \tilde{g}_0^2} \right)^2.
\end{align}
Thus, up to an overall factor corresponding to the function $h_{(1)}\left( \theta, \tilde{g}_0 \right)$ given by \eqref{xxz:h1}, the $S$-matrix \eqref{8v:Smatrix_solution1} amounts to the $XXZ$ model with interaction constants given by \eqref{xxz:JS1}.

\subsection{Solutions 2 and 3}

The analysis corresponding to solutions 2 and 3 requires that we consider the more general $S$-matrix describing the 8-vertex model on a square lattice. We recall that it can be parametrized in terms of elliptic functions as \cite{Baxter:1982zz,Samaj:2013yva,Khachatryan:2012wy}: 
\begin{equation} \label{xxz:Smatrix_xyz}
   S_{\textrm{(XYZ)}} =  \begin{pmatrix}
      \frac{\sn(u+\eta,k)}{\sn(u,k)} & 0 & 0 & k\,e^{\gamma/2}\sn(u+\eta,k)\sn(u,k) \\
        0 & \frac{\sn(u,k)}{\sn(\eta,k)} & 1 & 0 \\
       0 & 1 & \frac{\sn(u,k)}{\sn(\eta,k)} & 0 \\
	 k\,e^{-\gamma/2}\sn(u+\eta,k)\sn(u,k) & 0 & 0 & \frac{\sn(u+\eta,k)}{\sn(u,k)}
     \end{pmatrix}.
   \end{equation}
Here, $\sn(u,k)$ is the Jacobi elliptic function of argument $u$ and modulus $k$, $\eta$ and $\gamma$ are complex parameters. The coupling constants of the Hamiltonian describing the corresponding $XYZ$ model can be obtained from the following relations \cite{Samaj:2013yva}:
\begin{align} \label{xxz:J8}
	J_{x} = 1+k\,\sn^{2}(\eta,k), \quad J_{y} = 1-k\,\sn^{2}(\eta,k) \quad \text{and} \quad J_{z} = \cn(\eta,k)\dn(\eta,k).
\end{align}

To match the $S$-matrix of the $XYZ$ model \eqref{xxz:Smatrix_xyz} with the $S$-matrix \eqref{8v:Smatrix_solution2} describing solution 2: $\left(\tilde{g}_{1}=-\tilde{g}_0, \tilde{g}_{2}=0, \tilde{g}_{3}=0\right)$, we introduce, according to \eqref{xxz:factorization}, the function:
\begin{align}\label{xxz:h2}
	h_{(2)}(\theta,\tilde{g}_{0})=\frac{2\tilde{g}_{0}\coth \theta}{i + 2\tilde{g}_{0}\coth \theta}.
\end{align}
The corresponding $S$-matrix becomes:
\begin{equation}\label{xxz:Smatrix_solution2a}
   \tilde{S}_{\textrm{(2)}} = \begin{pmatrix}
      \frac{-1+2i\tilde{g}_{0}\coth \theta }{4\tilde{g}^{2}_{0} +2i\tilde{g}_{0} \coth \theta } & 0 & 0 & \frac{2\tilde{g}_{0}+i\tanh \theta }{2\tilde{g}_{0}+i\coth \theta } \\
        0 & \frac{i\tanh \theta }{2\tilde{g}_{0}} & 1 & 0 \\
       0 & 1 & \frac{i\tanh \theta }{2\tilde{g}_{0}} & 0 \\
	   \frac{2\tilde{g}_{0}+i\tanh \theta }{2\tilde{g}_{0}+i\coth \theta } & 0 & 0 &  \frac{-1+2i\tilde{g}_{0}\coth \theta}{4\tilde{g}^{2}_{0} +2i\tilde{g}_{0} \coth\theta}
     \end{pmatrix}.
   \end{equation}
Similarly, we can use the same function, 
\begin{align}\label{xxz:h3}
	h_{(3)}(\theta,\tilde{g}_{0}) = h_{(2)}(\theta,\tilde{g}_{0})
\end{align}
to factorize the $S$-matrix \eqref{8v:Smatrix_solution3} describing solution 3: $\left( \tilde{g}_{1}=0, \tilde{g}_{2}=-\tilde{g}_{0}, \tilde{g}_{3}=0 \right)$. The resulting $S$-matrix in this case is
\begin{equation}\label{xxz:Smatrix_solution3a}
   \tilde{S}_{\textrm{(3)}} = \begin{pmatrix}
      \frac{-1+2i\tilde{g}_{0}\coth \theta }{4\tilde{g}^{2}_{0} +2i\tilde{g}_{0} \coth \theta } & 0 & 0 & - \frac{2\tilde{g}_{0}+i\tanh \theta }{2\tilde{g}_{0}+i\coth \theta } \\
        0 & \frac{i\tanh \theta }{2\tilde{g}_{0}} & 1 & 0 \\
       0 & 1 & \frac{i\tanh \theta }{2\tilde{g}_{0}} & 0 \\
	   - \frac{2\tilde{g}_{0}+i\tanh \theta }{2\tilde{g}_{0}+i\coth \theta } & 0 & 0 &  \frac{-1+2i\tilde{g}_{0}\coth \theta}{4\tilde{g}^{2}_{0} +2i\tilde{g}_{0} \coth\theta}
     \end{pmatrix}.
   \end{equation}

We can, therefore, relate the $S$-matrix of the $XYZ$ model \eqref{xxz:Smatrix_xyz} and the above matrices $\tilde{S}_{\textrm{(2)}}$ and $\tilde{S}_{\textrm{(3)}}$ as follows. Setting the parameter $\gamma=0$ in \eqref{xxz:Smatrix_xyz}, identifying the variable $u$ in \eqref{xxz:Smatrix_xyz} with the rapidity $\theta$, \emph{i.e.}, making $u=\theta$, and taking the limit of the modulus $k \to 1$ one reproduces exactly the matrix in \eqref{xxz:Smatrix_solution2a} starting from the $S$-matrix of the $XYZ$ model, provided the following relation for the $\eta$ parameter:
\begin{align}\label{xxz:eta_relation}
	\eta=\cosh^{-1}\left[\frac{1}{\sqrt{1+4\tilde{g}^{2}_{0}}}\right]
\end{align}
is satisfied. To obtain the matrix in \eqref{xxz:Smatrix_solution3a}, one proceeds exactly as before but with $\gamma = 2 \pi i$. It then follows from \eqref{xxz:J8} that the $J_{x},J_{y},J_{z}$ interaction constants in the Hamiltonian for the $XYZ$ model take the form:
 \begin{align}\label{xxz:JS23}
 	J_{x} = 1 - 4 \tilde{g}_0^2 \quad \text{and} \quad J_{y}=J_{z} =1 + 4 \tilde{g}_0^2.
 \end{align}
Thus, up to an overall factor, the function $h_{(2)}(\theta,\tilde{g}_{0})$ in \eqref{xxz:h2} of rapidities and the coupling constant $\tilde{g}_{0}$, the $S$-matrices, $S_{\textrm{(2)}}$ \eqref{8v:Smatrix_solution2} and $S_{\textrm{(3)}}$ \eqref{8v:Smatrix_solution3}, correspond to the $XXZ$ model, with the interaction constants given by \eqref{xxz:JS23} depending on the coupling constant $\tilde{g}_{0}$ of the corresponding fermionic models \eqref{ext:explicit_Lagrangian_density_Sol2} and \eqref{ext:explicit_Lagrangian_density_Sol3}.

\subsection{Solution 4}
The $S$-matrix \eqref{8v:Smatrix_solution4} corresponding to solution 4: $\tilde{g}_{1,2,3}=\frac{1}{3}\left(-\tilde{g}_{0} \pm \sqrt{3+4\tilde{g}_{0}^{2}}\right)$ can be easily mapped to one of the exceptional solutions considered by \cite{Khachatryan:2012wy}:
\begin{align}\label{xxz:Smatrix_x-xz}
	S_{(X\_XZ)} = \begin{pmatrix}
		\sin v & 0 & 0 & \sin u\\
		0 & 0 & \sin (u + v) & 0\\
		0 & \sin (u+v) & 0 & 0\\
		\sin u & 0 & 0 & \sin v
	\end{pmatrix}.
\end{align}
It suffices to set $u=0$ and
\begin{align}
	v = - \sin^{-1} \left[ \frac{-2 \tilde{g}_{0} + \sqrt{3 + 4 \tilde{g}_0^2} \pm 3 i \coth \theta }{-2 \tilde{g}_{0} + \sqrt{3 + 4 \tilde{g}_0^2} \mp 3 i \coth \theta} \right]
\end{align}
in \eqref{xxz:Smatrix_x-xz} to obtain the $S$-matrix \eqref{8v:Smatrix_solution4}. This exceptional solution to the $YBE$ corresponding to the case $b_1 = b_2 = 0$ can be reduced to the usual $S$-matrix of the $XXZ$ model.

\section{Conclusion}
\label{section:Conclusion}
Motivated by finding a simpler theory with interaction terms similar to that of the $AAF$ model, we considered in this work the integrable properties of the Belavin model, and some of its anisotropic extensions. The $AAF$ model is a quite complex fermionic model which is hard to investigate using the standard methods in the context of integrable system, due to its highly non-ultalocal nature and singular potentials. On the other hand, the form of the $AAF$ action does resemble the typical $\nicefrac{1}{m}$ expansion arising in the  low-energy limit of some more fundamental theory. Thus, our motivation was to find a massive two-dimensional fermionic integrable model which had enough fermionic degrees of freedom to perform such a low-energy $\nicefrac{1}{m}$ expansion (see for example \cite{Fuentes-Martin:2016uol} for recent methods). 

We have found in this paper, for the simplest $su(2)$ invariant Belavin model, that the integrability in the massive case requires the investigation of exceptional solutions to the eight-vertex model. Furthermore, we showed that under some conditions on the coupling constants such exceptional solutions indeed exist. This is to contrast with the massless case, which had been extensively investigated, for which one can write Baxter's general solution for the eight-vertex model. It is easy to perform the low-energy expansion, at least to the lowest order, for one of our integrable solutions, \emph{e.g.}, for the interaction Lagrangian given in \eqref{ext:explicit_Lagrangian_density_Sol4a}. The resulting two-component massive fermionic model indeed has a form similar to that of the $AAF$ model with the characteristic $\nicefrac{1}{m}$ expansion (the explicit lengthy expression for the $AAF$ Hamiltonian can be found in \cite{Melikyan:2016gkd}). Although in this case we do not exactly reproduce the terms of the desired form, the resulting low-energy model does exhibits some similar features. Namely, as discussed in the introduction section, the interaction terms contain the derivatives of the fields, which would make the investigation of this model quite a difficult task, had we not had known that it is the low-energy expansion of a much simpler integrable model. There is of course still much work to do in order to reproduce exactly the terms of the $AAF$ model, and this problem remains open.

There are several ways to extend our results. Firstly, one can add more general interaction terms and investigate the existence of  exceptional solutions to the $YBE$ corresponding to the inhomogeneous form of the $S$-matrix. Secondly, it is straightforward to generalize our results to $su(n)$ case, and it would be interesting to find all non-trivial integrable models for the massive case, which would involve the exceptional solutions. Finally, we mention the following open problem, which we leave to a future publication. It was shown previously in \cite{Melikyan:2014yma,Melikyan:2016gkd} that the $AAF$ and the free fermion models are non-ultralocal. Since there is so far no lattice version of the $AAF$ model,  it is an interesting problem to first relate the lattice formulation of the free fermion model \cite{Bazhanov:1984iw,Bazhanov:1984ji,Bazhanov:1984jg,Burdik:2014ik} to the continuous limit, which results in a non-trivial Lax  pair and a non-ultralocal integrable structure.

\section*{Acknowledgments}  A.M. would like to thank Prof. Dr. Alvaro Ferraz - currently the director of the  \emph{International Institute of Physics, Natal}, whose exemplary  professionalism has motivated and inspired  this work.

G.W. would like to thank B. Cuadros-Melgar, E. Triboni, F. Florenzano and T. Lacerda for fostering a more open and vibrant scientific environment.

\phantomsection
\addcontentsline{toc}{section}{References}
\bibliographystyle{utphys}
\bibliography{Exceptional_Belavin_model_references}

\end{document}